\documentclass[defaultstyle,12pt]{thesis}

\usepackage{supertabular}
\usepackage{graphicx}
\usepackage{times}
\usepackage{color}
\usepackage[cmex10]{amsmath}
\usepackage{latexsym}
\usepackage{epsfig}
\usepackage{amsfonts}
\usepackage{amssymb}
\usepackage{url}
\usepackage{subfigure}
\usepackage{fancyhdr}
\usepackage{acronym}
\usepackage{algorithmic}
\usepackage{algorithm}
\usepackage{listings}
\usepackage{times}
\usepackage{cite}
\usepackage{array}
\usepackage{mdwmath}
\usepackage{mdwtab}
\usepackage{epsfig}
\usepackage{caption}
\usepackage{float}
\usepackage{pstricks}
\usepackage{multirow}

\newsavebox{\savepar}
\newenvironment{boxit}{\begin{lrbox}{\savepar}
                        \begin{minipage}[b]{6in}}
                         {\end{minipage}\end{lrbox}\fbox{\usebox{\savepar}}}

\title{Search for the PN coefficients for the Energy flux through Gravitational Waves from Black-Hole Binaries using Markov Chain Monte Carlo}

\author{Prayush}{Kumar}

\otherdegrees{}

\degree{Bachelor of Engineering (Honours)}{B.E.(Hons)}
\advisor{Dr}{Badri Krishnan}
\dept{Group of}{Astrophysical Relativity}

\abstract{
In this work, the focus is on the improvement of the existing post-Newtonian approximation for the gravitational flux from Super Massive Black Hole Binaries. In order to improve the existing templates for LISA, we need more accurate post-Newtonian expansions for the gravitational flux. Stochastic search techniques like the Markov Chain Monte Carlo (MCMC) have been used extensively for searching for sky parameters etc. The idea is to combine the two and approach the problem of finding post-Newtonian coefficients using MCMC. It has been shown that matching against a 5.5PN signal, with noise, the last coefficient can be found by MCMC very easily and displays fast convergence. Also the space for higher dimensional searches are explored. 
}

\acknowledgements{
First and foremost, I would like to express my gratitude to my supervisor Dr Badri Krishnan, for his constant guidance, instructions, encouragement and occasional timely reminders, which were instrumental in the completion of this thesis. His support has been extremely valuable.

Then, I wish to sincerely thank Dr Edward Porter for his insight during the frequent discussions that we had in the development of the ideas espoused in this thesis. His expertise played a pivotal role in shaping the ideas presented in this thesis.

Also, I wish to thank my friend, Srikrishna Sridhar, who helped me in a number of ways\cite{Latex}. 

I would also like to express my great appreciation to the Max Planck Institute for Gravitational Physics, Potsdam; and Birla Institute of Technology and Science (BITS) for providing all the resources and pleasant atmosphere for conducting research. Without their support, I would not have been able to embark on this.
}
\begin{document}
\chapter{Introduction}
\label{ch:Introduction}

\section{Background}
Mass in nonspherical and nonuniform motion is the source of ripples in curved spacetime, which propagate away at the speed of light. These propagating ripples in spacetime curvature are called \textit{gravitational waves} (GW). The weak coupling to matter is what makes GW so difficult to detect, and so astrophysically interesting. Once produced, little is absorbed, unlike electromagnetic radiation. GW, in principle, could enable us to see closer to the horizon of a black hole and to earlier moments in the universe than with any form of electromagnetic radiation.

Laser Interferometer Space Antenna (LISA), a joint venture of ESA and NASA, would be capable of detecting gravitational waves in the frequency band of $10^{-5}Hz < f < 1$ Hz. Promising a future direct observation of gravitational radiation, it would test one of the most fascinating predictions of General Relativity, and, at the same time, becoming a new powerful tool in the astronomical investigation of highly relativistic catastrophic events, such as the merging of compact binary systems and the collapse of massive stellar cores. The inspiral and merger of massive black hole binaries (MBHBs) are a source of the strongest gravitational wave (GW) signals in LISA's frequency band, specifically in the sub mHz range. This frequency band, is not observable by the ground based detectors, because of a few reasons, such as the interference from the tectonic motion below around 10Hz, and, the time varying gravitational potentials such as the weather systems etc.

The large signal-to-noise ratios (SNRs) of these signals will allow LISA to not only detect, but measure physical parameters of the source systems.  This would hold immense potential to probe questions regarding the role played by black holes in structure formation and galactic dynamics. Estimates of the precision with which LISA can extract waveform parameters have been developing for the past several years, in parallel with progress in describing MBHB waveforms. The parameters are extracted, in essence, by cross - correlating the signal (with noise) with theoretical templates. The waveforms from inspiraling binaries are calculated using the post-Newtonian expansion of Einstein equations, which assumes that the system comprises of slow moving bodies. The post Newtonian approximations, however, break down at or near the merger and are not expected to give accurate description of the merger waveform. So I rely on the part of the signal that extends till some time before that. Also, if the template and the signal lose phase with each other by even one cycle, their cross correlation significantly reduces. This means, that, the theoretical templates would have to be better than one cycle during an entire sweep through LISA's band. Althought the post-Newtonian calculation technique is being developed to apply to higher order calculation, it would be productive if there were a faster and accurate way to obtain the higher order post-Newtonian corrections.

In this thesis, I look at a system of non-spinning extreme mass ratio SMBHB inspirals, and the idea is to let stochastic search methods like Markov Chain Monte Carlo (MCMC) to search for the post-Newtonian coefficients beyond the 5.5PN. We use the form of the function as in \cite{PN55}, and also aim to validate the same.

\section{Motivation}
This thesis, is the summary of the work I did and things I read during my visit at the Max Planck Institut f\"{u}r Gravitationsphysik, Potsdam. The guidance of Dr Badri Krishnan, and Dr Edward Porter was extremely valuable, in terms of giving direction to my learning. 

Stochastic search methods, like the Markov Chain Monte Carlo, have been used for estimation of parameters describing various gravitational waves' sources, which are going to be observed by LISA \cite{LISA}. This is done by matching the observed signal against the pre-evolved theoretical templates. The stochastic search is used in exploring the massive banks of these templates. Extremely accurate post-Newtonian waveforms to create these template banks. 

The inspiration of the thesis crystallized around the idea of letting stochastic search find the higher order coefficients to the existing post-Newtonian forms. And in seeing, for instance, that with the knowledge of theoretically derived post-Newtonian coefficients, if the Markov Chains could find the higher order coefficients.

\section{Thesis Organization}

The thesis is organized into six chapters.

The first being the Introductory chapter, the second talks briefly about the General Theory of Relativity, the meaning of curvature, Riemann tensor, the Einstein's Equations and also covers the Schwarzschild Geometry, Gravitational waves, the Transverse-Traceless gauge simplfication. A brief derivation of Newtonian gravitational flux, from the quadrupole moment formula is also presented.

The third chapter, would focus on describing the mathematical modelling of the LASER Inerferometer Space Antenna (LISA), and the form of the gravitational waveform as observed by it.

The fourth chapter, focusses on the Markov Chain Montel Carlo method, which is the primary algorithm I had chosen for all the searches. Here, I would illustrate the main search technique. This would comprise of an introduction of the method, followed by the most of the 'how' of what I did.

The fifth chapter would state the results obtained, supported with some explanation, and the sixth chapter would present the Conclusion, and the scope for future work on this.
\chapter{Gravitational Waves\cite{Maggiore}\cite{GW0}\cite{GW2}}

A \textit{black-hole} is a region in spacetime in which the gravitational field is so strong that it doesn't allow even light to escape to infinity. It is formed when a body of mass $M$ contracts to a size less than, what is commonly refered to as, the \textit{gravitational radius $r_g = 2GM/c^2$}. The velocity required to leave the boundary of the black-hole and escape equals the speed of light. As the speed of light is the limiting propagation velocity for physical signals, its obvious that absolutely nothing can escape from the region inside the \textit{black} body.

In a short time following its formation, a black-hole becomes stationary and its field can be uniquely described with the knowledge of its mass, angular momentum and its electric charge (if charged). This is because, in the extremely strong field of the black-hole, only very special configurations of physical fields (including gravitational field) can be stationary. Einstein's equations give the description of the field around the black-hole. 

Since no signals can escape a black-hole, but physical objects and light can fall into it, the spacetime surface (event horizon) of a black-hole is light-like. From this it follows that processes involving a black-hole would be irreversible. If it remains isolated, and eventually attains a stationary state. To understand the stationary geometry, I would like to discuss certain more basic things in brief.

\section{Curvature in spacetime}
A line element specifies the geometry of spacetime. Its expression depends on the coordinate system employed, but the essence remains the same. 

For instance, flat spacetime geometry can be described, in Cartesian coordinate system, by:
\begin{equation}
 ds^2 = -c^2 dt^2 + dx^2 + dy^2 + dz^2			\nonumber
\end{equation}
or in polar coordinate system as:
\begin{equation}
 ds^2 = -c^2 dt^2 + dr^2 + r^2 d\theta^2 r^2 \sin^2\theta d\phi^2		\nonumber
\end{equation}
In general, a \textit{metric} is used to define the geometry. If $x^{\alpha}$ represents the points in spacetime, the line element joining nearby points, has its length $ds$ given by the following expression:
\begin{equation}\label{eq:metric}
 ds^2 = g_{\alpha\beta}(x) dx^{\alpha}dx^{\beta}
\end{equation}
In this fashion, the flat space time, would have its metric as ($dx^0 = cdt$) $diag(-1,1,1,1)$ in Cartesian coordinates, and $diag(-1,1,r^2,r^2\sin^2\theta)$ in polar. This is also known as the \textit{Minkowski} metric of flat spacetime, and typically represented by $\eta_{\alpha\beta}$ 

Einstein's equivalence principle states that:
\linebreak
\begin{boxit}
All test particles at the same spacetime point in a given gravitational field will undergo the same acceleration, independent of their properties, including their rest mass, and, The outcome of any local non-gravitational experiment in a laboratory moving in an inertial frame of reference is independent of the velocity of the laboratory, or its location in spacetime.
\end{boxit}

Here \textit{local} has a very special meaning: not only must the experiment not look outside the laboratory, but it must also be small compared to \textit{variations} in the gravitational field, tidal forces, so that the entire laboratory is moving inertially.

This implies that the local properties of curved spacetime would be essentially Minkowskian. This is an extremely important concept, as it implies that at any point in spacetime in curved spacetime (described by the metric $g_{\alpha\beta}(x)$), we can always introduce a new system of coordinates $x^{'\alpha}$ such that:
\begin{equation}\label{eq:localinertial}
 g^{'}_{\alpha\beta}(x^{'}) = \eta_{\alpha\beta}
\end{equation}

Also, in curved spacetime, \textit{vectors} is a concept that is defined \textit{locally}. It is a concept that was used in the flat spacetime, and since the equivalence principle suggests that the local properties of curved spacetime are Minkowskian, it becomes a concept defined locally at each point. This means, that at a point, normal vector operations are valid, but between vectors defined at different spacetime points, they are not. For that, they have to be first parallel-transported to the same point first. This means, that vectors at different points cannot be added or compared, just like that. Vectors are expressed in terms of locally defined coordinate bases. The two most commonly used bases are \textit{the orthonormal bases} and \textit{the coordinate bases}. In the latter, the unit vectors are such that:
\begin{equation}\label{eq:coordinatebases}
 \textit{e}_{\alpha}(x)\textit{e}_{\beta}(x) = g_{\alpha\beta}(x)
\end{equation}

\subsection{Geodesics}
How a test particle or a light ray moves in a curved spacetime, would tell us how the \textit{curved} spacetime is \textit{curved}. If a test mass is introduced into the physical scenario, it would inevitably move according to the curvature produced by other bodies with significant masses. This path is called a \textit{geodesic}. In every local Lorentz frame, this curve would appear to be straight and uniformly parameterized. In other words, a geodesic is a curve $C(\lambda)$, that parallel-transports its tangent vector $\textit{u} = \frac{dC}{d\lambda}$ along itself. 
ie:
\begin{equation}
\nabla_u u = 0
\end{equation}

And it leads to the \emph{geodesic equation} :

\begin{equation}\label{eq:geodesiceqn}
\frac{d^2 x^{\alpha}}{d\lambda^2} + \Gamma^{\alpha}_{\mu\gamma} \frac{dx^{\mu}}{d\lambda}\frac{dx^{\gamma}}{d\lambda} = 0
\end{equation}

The $\Gamma^{\alpha}_{\mu\gamma}$ are the connection coefficients, also known as the \textit{Christoffel symbols}, and $\lambda$ is called the \textit{affine parameter} (parameterizing the geodesic).

A few properties of the Christoffel symbols are important, that they can be taken as symmetric in the lower two indices. ie:
\begin{equation}\label{eq:christoffelsymmetry}
 \Gamma^{\alpha}_{\beta\gamma} = \Gamma^{\alpha}_{\gamma\beta}
\end{equation}

and its expression in terms of the general metric (and its derivatives) is:
\begin{equation}\label{eq:christoffelandmetric}
g_{\alpha\delta} \Gamma^{\delta}_{\beta\gamma} = \frac{1}{2} \big( \frac{\partial g_{\alpha\beta}}{\partial x^{\gamma}} + \frac{\partial g_{\alpha\gamma}}{\partial x^{\beta}} - \frac{\partial g_{\beta\gamma}}{\partial x^{\alpha}} \big)
\end{equation}

This expression, along with various symmetries of the spacetime, can be easily juggled with, to evaluate the expressions for the Christoffel symbols. Given an initial location in spacetime, and an initial four-velocity, the geodesic equation can be easily integrated numerically to find the location and four-velocity at later moments of proper time. An important expression, employed to reduce the order and the number of equations is that of taking the four-vector dot product of velocity with itself.
\begin{equation*}\label{eq:udotu}
 \textit{u}\cdot\textit{u} = g_{\alpha\beta} \frac{dx^{\alpha}}{d\tau} \frac{dx^{\beta}}{d\tau} = -1
\end{equation*}

\subsection{Riemann Curvature}
The motion of two test particles is enough to detect spacetime curvature. By studying the \textit{relative} motion of two \textit{nearby} test particles, can a person detect the local curvature of the spacetime. By \textit{nearby}, I mean particles traveling on infinitesimally seperated geodesics. Let four-vector $\vec{\chi}$ denote the infinitesimal displacement between two \textit{nearby} geodesics. The expression for $\nabla_u \nabla_u \chi$ would give the acceleration of the seperation vector. Which would give the local curvature of spacetime
\cite{Weinberg-Riemann} states this relation as:
\begin{equation}
 (\nabla_u \nabla_u \chi)^\alpha = - R^{\quad\alpha}_{\beta\gamma\delta} u^{\beta}\chi^{\gamma}u^{\delta}
\end{equation}
which leads to the expression for the Riemann curvature tensor:
\begin{equation}\label{eq:riemanncurvature}
\fbox{$R^{\alpha}_{\quad\beta\gamma\delta} = \frac{\displaystyle\partial \Gamma^{\alpha}_{\beta\delta}}{\displaystyle\partial x^{\gamma}} - \frac{\displaystyle \partial \Gamma^{\alpha}_{\beta\gamma}}{\displaystyle\partial x^{\delta}} + \Gamma^{\alpha}_{\gamma\epsilon}\Gamma^{\epsilon}_{\beta\delta} - \Gamma^{\alpha}_{\delta\epsilon}\Gamma^{\epsilon}_{\beta\gamma}$}
 \end{equation}

Often, its fully covariant form $R_{\alpha\beta\gamma\delta} ( = g_{\alpha\sigma} R^{\sigma}_{\quad\beta\gamma\delta})$ is also refered to as the Riemann curvature tensor.

The curvature of spacetime at each point is completely described by this \textit{multilinear operator}, which has 20 algebraically independent components at each point. The components of the Riemann tensor identically satisfy a differential equation (the Bianchi identities\cite{Weinberg-Bianchi}) and certain symmetries\ref{eq:Riemann-symmetries}, which is why the metric tensor\ref{eq:metric} (which has ten algebraically independent components at each pointt) is enough to completely determine the Riemann curvature tensor.
\begin{eqnarray}\label{eq:Riemann-symmetries}
R_{\alpha\beta\gamma\delta} &= - R_{\beta\alpha\gamma\delta} \\ \nonumber
R_{\alpha\beta\gamma\delta} &= - R_{\alpha\beta\delta\gamma} \\ \nonumber
R_{\alpha\beta\gamma\delta} &= + R_{\gamma\delta\alpha\beta} \\
R_{\alpha\beta\gamma\delta} + R_{\alpha\delta\beta\gamma} + R_{\alpha\gamma\delta\beta} &= 0 \nonumber
\end{eqnarray}

For instance, in the local inertial frame, where $g_{\alpha\beta} = \eta_{\alpha\beta}$, the Riemann curvature simplifies to:

\begin{equation}\label{eq:Riemann:g=eta}
 R_{\alpha\beta\gamma\delta} = \frac{1}{2} \big( \frac{\partial^2 g_{\alpha\delta}}{\partial x^{\beta} \partial x^{\gamma}} - \frac{\partial^2 g_{\alpha\gamma}}{\partial x^{\beta} \partial x^{\delta}} - \frac{\partial^2 g_{\delta\beta}}{\partial x^{\alpha} \partial x^{\gamma}} + \frac{\partial^2 g_{\beta\gamma}}{\partial x^{\alpha} \partial x^{\delta}}\big)
\end{equation}
Another useful tensor, the \textit{Ricci curvature tensor}, is defined as:
\begin{equation}
 R_{\mu\nu} = R^{\alpha}_{\quad\mu\alpha\nu}
\end{equation}
\pagebreak

\subsection{Einstein's Equations}\label{subsec:EinsteinEqns}
Finally, I am in a position to introduce the Einstein tensor and the Einstein equation. As argued in MTW, the frame independent stress-energy tensor \emph{T} must act as the source of gravity. To link it with the geometry of spacetime, \emph{T} would need to be expressed in terms of a tensor that describes in some way, the geometry of spacetime which is a result of gravity.

This is the Einstein tensor \emph{G} that we are talking about, and it must be that:
\begin{enumerate}
\item \emph{G} vanishes when spacetime is flat.
\item \emph{G} is constructed from the Riemann curvature tensor \ref{eq:riemanncurvature} and the metric \ref{eq:metric} only.
\item \emph{G} has to be linear in \emph{R}, to be substantiated as a measure of curvature.
\item \emph{G} must be symmetric, second rank, have a zero divergence.
\end{enumerate}

Finally, \emph{G} is defined as (with due justification):

\begin{equation}\label{eq:defineG}
 G_{\mu\nu} = R_{\mu\nu} - \frac{1}{2}g_{\mu\nu}R
\end{equation}
where $R = g^{\mu\kappa} R_{\mu\kappa}$ is the \textit{curvature scalar}. And, this leads to the Einstein field equations:
\begin{equation}\label{eq:EinsteinEqn}
 \emph{G} = 8\pi \emph{T}
\end{equation}

The Riemann tensor $R_{\alpha\beta\gamma\delta}$ can be decomposed into two pieces, the Ricci tensor and the Weyl tensor $C_{\alpha\beta\gamma\delta}$, in a manner analogous to decomposing a matrix into trace and tracefree parts. The Riemann, Ricci, and Weyl tensors all have geometric meaning independent of any physical interpretation.

Physical meaning enters via the stress-energy tensor \emph{T} which can be thought of as a 4x4 symmetric matrix (so it has 10 algebraically independent components at each point). This tensor completely describes the amount of (non-gravitational) mass-energy at each point, and also any momentum (mass-energy flow) and stresses (such as the pressures in a fluid).

The Ricci curvature is directly coupled to the immediate presence of matter at a given point. If there is no mass-energy at a given point, the Ricci tensor vanishes.

\section{Spacetime around a black-hole: the Schwarzschild \\Geometry}
The Schwarzschild geometry describes the spacetime geometry of empty space surrounding any spherical mass. Karl Schwarzschild derived this geometry in 1915, within a few weeks of Albert Einstein publishing his fundamental paper on the Theory of General Relativity. Due to the symmetries, this is a  case when analytical solution to the Einstein field equation can be obtained, and this geometry is precisely the geometry given by the solution of Einstein equations in vacuum.

As stated earlier, the line element completely represents the geometry. The line element summarizing the Schwarzschild geometry is given by:
\begin{equation*}\label{eq:schwarzschildmetric}
\fbox{$
ds^2~=~-\big( 1-\frac{\displaystyle 2GM}{\displaystyle c^2r}\big)(cdt)^2~+~\big( 1-\frac{\displaystyle 2GM}{\displaystyle c^2r}\big)^{-1}(cdt)^2~+~r^2\big(d\theta^2+\sin^2\theta d\phi^2\big)
$}
\end{equation*}
The coordinates are called \textit{Schwarzschild} coordinates, and the corresponding $g_{\alpha\beta}$ is called the \textit{Schwarzschild} metric. The important properties of the metric are that it is \textit{time independent} and \textit{spherically symmetric}. Also, it is determined by a single parameter $M$, which is the total mass of the gravitational source which produces the field.

The proper time, is given by $d\tau = \sqrt{-g_{00}} = \big(1-\frac{\displaystyle 2GM}{\displaystyle c^2 r}\big)^\frac{1}{2}dt$. 

Also, the coordinate $r$ is not physically the distance from any origin. Like, as $r \rightarrow \infty$, the proper time $\rightarrow dt$, ie the geometry tends to become Minkowskian. But, as $r$ becomes progressively smaller and approaches the value $\frac{\displaystyle 2GM}{\displaystyle c^2}$, the proper time interval decreases. As the Schwarszchild coordinate $r$ comes closer to $\frac{\displaystyle 2GM}{\displaystyle c^2}$, the time in the local Lorentz frames passes very very slowly. This value is called the \textit{Schwarzschild radius}.
\begin{equation*}\label{eq:r_g}
 r_g = \frac{2GM}{c^2}
\end{equation*}

In fact, Hartle\cite{Hartle} says: [..It ($r$) is related to the area A of the two-dimensional spheres of fixed radius $r$ and $t$ by the standard formula $r = (A/4\pi)^{1/2}$..]

\subsection{Trajectories in Schwarzschild spacetime: the LSO}
For schwarzschild spacetime, $g_{\alpha\beta}$ is independent of time, and of $\phi$.

For a test-mass, $E = -p_0/m$, and $L = p_{\phi} / m$. This means $p_0$ and $p_{\phi}$ would be constant for a trajectory. So would $E$ and $L$ be. 
Now, for a particle:
\begin{eqnarray}
p^0 &=& g^{00} p_0 = m\big(1-\frac{\displaystyle 2GM}{\displaystyle r}\big)^{-1} E \\ \nonumber
p^r &=& m dr/d\tau \\
p^{\theta} &=& 0 (\text{fixing the trajectory plane}) \\ \nonumber
p^{\phi} &=& g^{\phi\phi}p_{\phi} = \frac{\displaystyle m}{\displaystyle r^2}L\\ \nonumber
\end{eqnarray}
Using the equation $\vec{p} \dot \vec{p} = -m^2$ with the above values yields:
\begin{equation}
 \big(\frac{dr}{d\tau}\big)^2 = E^2 - \big(1-\frac{2GM}{r}\big)\big(1+\frac{L^2}{r^2}\big)
\end{equation}

which translates to:
\begin{equation}\label{eq:totalELSO}
E^2 = \big(\frac{dr}{d\tau}\big)^2 + \big(1-\frac{2GM}{r}\big)\big(1+\frac{L^2}{r^2}\big)
\end{equation}

giving the effective potentials as:
\begin{equation}
V^2(r) = \big(\frac{dr}{d\tau}\big)^2 + \big(1-\frac{2GM}{r}\big)\big(1+\frac{L^2}{r^2}\big)
\end{equation}

The total $E$ must remain constant. So, \ref{eq:totalELSO} can be differentiated with respect to the coordinate $r$, which will give the relation:
\begin{equation}
\frac{d^2r}{d\tau^2} = \frac{1}{2}\frac{d}{dr}V^2(r)
\end{equation}

\begin{figure}[h]
\centering
\includegraphics[keepaspectratio=true,width=5.5in]{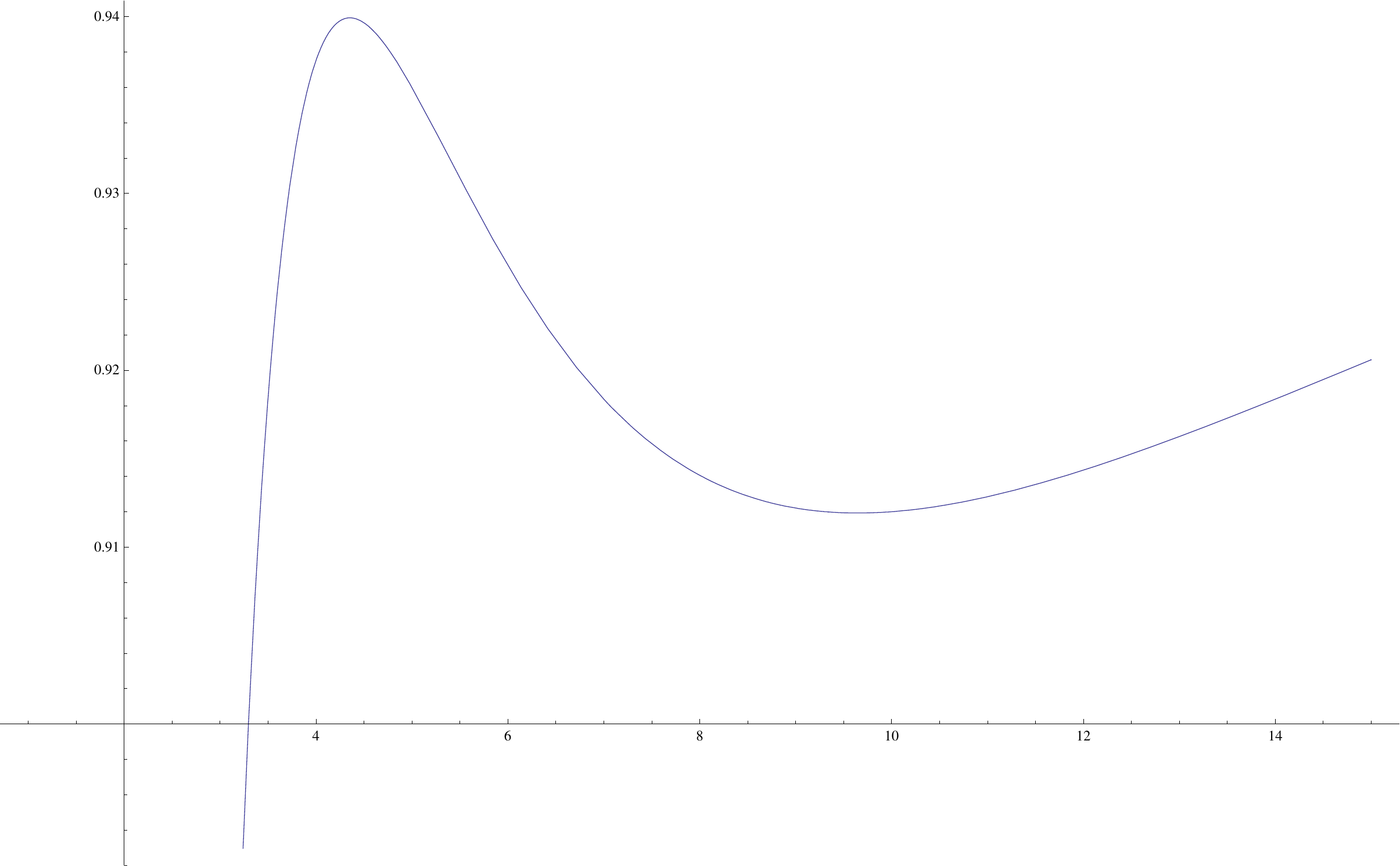}
\caption{Typical effective potential for a massive particle of fixed specific angular momentum in the Schwarzschild metric.}
\label{fig:V2}
\end{figure}

This physically means, that an orbit in this geometry can be stable only at an extremum of $V^2(r)$, as in Fig. \ref{fig:V2}. Putting $\frac{d}{dr}V^2(r) = 0$, and using \ref{eq:totalELSO}, 
\begin{equation}
\frac{\displaystyle d}{\displaystyle dr}\big[\big(1-\frac{2GM}{r}\big)\big(1+\frac{L^2}{r^2}\big)\big] = 0
\end{equation}

Expressing in terms of geometrized units, where $c = 1~\&~G = 1 (dimensionless)$, this operation leads to:
\begin{equation}
r = \frac{L^2}{2M}\big(1\pm \surd\big(1-\frac{12M^2}{L^2}\big)\big)
\end{equation}

\textit{Note: I will be using geometrized units henceforth.)}

This means, two values of $r$ exist for a given constant $L$ (only when $L \geq 12M^2$). If, however, the $L$ is lower than the minimum value of $\sqrt{12}M$, the particly does insufficient angular momentum to stabilize at any orbit, and continues on into the black-hole and plunges.

As there is a minimum value of $L$ for a particle in stable circular orbit, there has to be a minimum $r$ also. This value of $r$ is know as the \textit{Last Stable Orbit} (LSO).
\begin{equation}\label{LSO}
 r_{LSO} = 6M
\end{equation}

The period for a particle in a stable circular orbit around a black-hole would have a time-period of $2\pi\surd(r^3/M)$. This expression, coincidentally matches with the Newtonian expression. 

\section{the Post Newtonian Approximation}
Post-Newtonian expansions in general relativity are used for finding an approximate solution of the Einstein equations for the metric tensor that represents a multi-component, tensor gravitational field potential instead of a single, scalar gravitational potential in the Newtonian gravity. In the limit, when the velocities involved are small as compared to $c$, the post-Newtonian expansions degenerate and the Einstein theory of general relativity is reduced to the Newtonian-like theory of gravity with the instantaneous action-at-the-distance gravitational field interaction.

Consider a system of particles that are bound together by the mutual gravitational forces. The post-Newtonian approximation may be described as a method for obtaining the motions of the system to higher powers of the parameters $G\bar{M}/\bar{r}$ and $\bar{v}^2$, than given by Newtonian mechanics. It is sometimes referred to as an expansion in inverse powers of the speed of light. But I'll take $c$ as unity. So, it becomes an expansion in $\bar{v}^2$.

The equation of motion for a particle is:
\begin{equation}
\frac{d^2 x^{\mu}}{d\tau^2} + \Gamma^{\mu}_{\nu\lambda} \frac{dx^{\nu}}{d\tau}\frac{dx^{\lambda}}{d\tau}
\end{equation}

From this the acceleration becomes:
\begin{eqnarray*}
\frac{d^2 x^i}{dt^2} &=& \big(\frac{dt}{d\tau}\big)^{-1} \frac{d}{d\tau} \big[ \big(\frac{dt}{d\tau}\big)^{-1} \frac{dx^i}{d\tau} \big]\\
                     &=& -\Gamma^i_{\nu\lambda} \frac{dx^{\nu}}{dt} \frac{dx^{\lambda}}{dt} + \Gamma^0_{\nu\lambda}\frac{dx^{\nu}}{dt}\frac{dx^{\lambda}}{dt}\frac{dx^i}{dt}\\
\end{eqnarray*}

Writing it explicitly, in a way where we would be able to make certain approximations more evidently;
\begin{equation}
\frac{d^2 x^i}{dt^2} = -\Gamma^i_{00} - 2\Gamma^i_{0j}\frac{dx^i}{dt} - \Gamma^i_{jk}\frac{dx^i}{dt}\frac{dx^k}{dt}\\
+\big[ \Gamma^0_{00} + 2\Gamma^0_{0j}\frac{dx^j}{dt} +\Gamma^0_{jk}\frac{dx^i}{dt}\frac{dx^k}{dt} \big] \frac{dx^i}{dt}
\end{equation}

What we need is a systematic approximation method, that will not rely on any assumed symmetry property of the system. One important such method is the \textit{post-Newtonian approximation}\cite{PN-Einstein}. It is adapted to a system of slowly moving particles, bound together by gravitational forces. I would try to briefly demonstrate how this systematic assumption works.

In the case discussed above, the objective is to compute $\frac{\displaystyle d^2 x^i}{\displaystyle dt^2}$ to order $\frac{\displaystyle \bar{v}^n}{\displaystyle \bar{r}}$. So the various components of the affine connection would be needed to the following orders.

\begin{eqnarray*}
\Gamma^i_{00} &\text{to order}& \frac{\bar{v}^n}{\bar{r}}\\
\Gamma^i_{0j} \& \Gamma^0_{00} &\text{to order}& \frac{\bar{v}^{n-1}}{\bar{r}}\\
\Gamma^i_{jk} \& \Gamma^0_{0j} &\text{to order}& \frac{\bar{v}^{n-2}}{\bar{r}}\\
\Gamma^0_{jk} &\text{to order}& \frac{\bar{v}^{n-3}}{\bar{r}}\\
\end{eqnarray*}

It is always possible to find a coordinate system in which the metric tensor is nearly equal to the Minkowski tensor\cite{Weinberg-g=h}, and the corrections are expandable in powers of $G\bar{M}/\bar{r} ~\approx~\bar{v}^2$. This gives:
\begin{eqnarray*}
g_{00} = -1 + g^2_{00} + g^4_{00} + \dots \\
g_{ij} = \delta_{ij} + g^2_{ij} + g^4_{ij} + \dots \\
g_{i0} = g^3_{i0} + g^5_{i0} + \dots
\end{eqnarray*}

where, $g^n_{ij}~denotes~g_{ij}~to~order~n.$

Using these approximations, the components $\Gamma^i_{00}, \Gamma^i_{jk}, \Gamma^0_{0i}$ can be expanded as:
\begin{equation}
\Gamma^{\mu}_{\nu\lambda} = \Gamma^{(2)\mu}_{\nu\lambda} + \Gamma^{(4)\mu}_{\nu\lambda} + \dots
\end{equation}

and, the components $\Gamma^i_{0j}, \Gamma^0_{00}, \Gamma^0_{ij}$ can be expanded as:
\begin{equation}
\Gamma^{\mu}_{\nu\lambda} = \Gamma^{(3)\mu}_{\nu\lambda} + \Gamma^{(5)\mu}_{\nu\lambda} + \dots
\end{equation}

Thus working, we'd be able to get the order to which the metric is to be evaluated, as the affine connection (Christoffel symbols) can be calculated from the metric alone [\ref{eq:christoffelandmetric}]. So can the Ricci tensor.

\section{In the Transverse - Traceless gauge}
Let the trace-reverse form of $h$ be defined as:
\begin{equation*}
\bar{h}^{\alpha\beta} = h^{\alpha\beta} - \frac{1}{2}\eta^{\alpha\beta}h.
\end{equation*}
Its called the trace-reverse form because its trace is equal to the negative of the trace of $h$. The weak-field approximation ($|h_{\alpha\beta}| \ll 1$) to gravitational field, leads to the weak-field Einstein equations:
\begin{equation}\label{eq:weakfield-Einsteineq}
\big( -\frac{\partial^2}{\partial t^2} + \nabla^2 \big) \bar{h}^{\alpha\beta} = -16\pi T^{\alpha\beta}
\end{equation}
For vacuum ($T^{\alpha\beta} = 0$), this has a known solution of the form:
\begin{equation}\label{eq:sol-weakfield-Einsteineq}
\bar{h}^{\alpha\beta} = A^{\alpha\beta} e^{ik_{\alpha}x^{\alpha}}
\end{equation}

As the coordinate system that is chosen is arbitrary, its extremely simplifying to impose certain restrictions on them. For approximate solutions to the weak-field Einstein's equations [\ref{eq:weakfield-Einsteineq}], the gauge can be changed using a vector that satisfies the following equation and still not affect anything.
\begin{equation*}
\big( -\frac{\partial^2}{\partial t^2} + \nabla^2 \big) \xi_{\alpha} = 0
\end{equation*}
A known solution to the equation is: $\xi_{\alpha} = B_{\alpha} e^{ik_{\mu}x^{\mu}}$, where $B_{\alpha}$ is a constant and $k^{\mu}$ is the same as in [\ref{eq:sol-weakfield-Einsteineq}]. This produces a change in $\bar{h}^{\alpha\beta}$, given by:
\begin{equation}\label{eq:hbar-gauge-transformation}
\bar{h}_{\alpha\beta}^{(NEW)} = \bar{h}_{\alpha\beta}^{(OLD)} - \xi_{\alpha,\beta} - \xi_{\beta,\alpha} + \eta_{\alpha\beta}\xi^{\mu}_{,\mu} 
\end{equation}
and for the [Eq \ref{eq:sol-weakfield-Einsteineq}]'s constant changes as:
\begin{equation*}
A_{\alpha\beta}^{(NEW)} = A_{\alpha\beta}^{(OLD)} - iB_{\alpha}k_{\beta} - iB_{\beta}k_{\alpha} + i\eta_{\alpha\beta}B^{\mu}k_{\mu}
\end{equation*}

Now, if the following conditions are imposed to the selection of the coordinates, immense simplification is attained:
\begin{eqnarray}\label{eq:TT}
A^{\alpha}_{\; \alpha} = 0 \nonumber \\ 
A^{\alpha\beta}k_{\beta} = 0 \\
A_{\alpha\beta}U^{\beta} = 0 \nonumber
\end{eqnarray}
where $\vec{U}$ is an arbitrary constant timelike unit vector. Of course, these conditions are imposed alongwith the basic Lorentz - gauge condition ($\bar{h}^{\mu\nu}_{\;\;,\nu} = 0$). These conditions are called the \textit{Transverse - Traceless (TT)} gauge conditions. \textit{Traceless} because the trace of $A$ vanishes.  \textit{Transverse} because $A_{\mu\nu}$ is 'across' the direction of propagation. The simplification that is attained here, is that in this frame:
\begin{equation*}
A^{TT}_{\alpha\beta}~\text{has only} A^{TT}_{xx},~A^{TT}_{xy} \neq 0.
\end{equation*}
All the remaining components vanish.! Thus there are only two independent components. The easiest way to write them explicitly is to orient the spatial coordinates so that one axis is  along the propagation of the waves. The $A^{TT}_{xx}$ and $A^{TT}_{xy}$ represent the two polarizations of the gravitational wave. The former is usually called the $+$ (plus) polarization, and the latter is referred to as the $\times$ (cross) polarization. The general solution of the weak-field equations is a superposition of the two polarizations.

To transform to the TT gauge, there are very simple relations which essenially set all the nontransverse parts of the metric equal to zero, and subtract out the trace from the remaining diagonal elements to make it traceless. For example, for a wave propagating in the z- direction,
\begin{equation*}
h^{TT}_{xx} = -h^{TT}_{yy} = \frac{1}{2} (h_{xx} - h_{yy}); ~ h^{TT}_{xy} = h_{xy}.
\end{equation*}

\section{the Newtonian Flux\cite{GW1} \cite{GW0}}\label{sec:NewtonianFlux}
The most simplistic representation of a system emitting gravitational waves can be through its quadrupole moment. The origin of this name rests in the analogy with electromagnetism. The electromagnetic field is a vector field, and so electromagnetic waves can be generated by vector sources, such as an electric current. This means that a dipole source is sufficient (a dipole can be described by a vector). Gravity, on the other hand, is a tensor field, and the source must contain more components than a dipole (vector) to simulate it. A tensor can be regarded as a conjunction of two vectors\cite{Feynman}, so the source must be at least as complicated as two vectors. The simplest such arrangement is the quadrupole, consisting of two opposed vector dipoles. The resulting field pattern will reflect this more complicated arrangement ($\sin ^2 \theta$ rather than $\sin \theta$ angular field dependence).

The quadrupole moment tensor of the mass distribution is given by:
\begin{equation}\label{eq:quadrupolemoment}
I^{lm} = \int T^{00} x^l x^m d^3 x
\end{equation}
and its trace - free  representation is given by $I^T_{jk} = I_{jk} - \frac{1}{3}\delta_{jk}I^l_{\; l}$

In the TT gauge, the solution to Equation \ref{eq:weakfield-Einsteineq} in terms of this quadrupole moment is given by:
\begin{eqnarray}\label{eq:quadpolemom-solution}
\bar{h}^{TT}_{xx} &=& \frac{1}{r} [I^T_{xx,00} (t-r) - I^T_{yy,00}(t-r)] \\
\bar{h}^{TT}_{xy} &=& \frac{2}{r} I^T_{xy,00}(t-r) \nonumber
\end{eqnarray}

For an isolated system which is emitting gravitational waves, with $\Omega$ is the frequency of the oscillations of the time varying part of the $T_{\mu\nu}$ tensor (assumed sinusoidal), its net gravitational flux at a distance r along the z axis is given as:
\begin{equation}
F = \frac{\Omega^6}{16\pi r^2} \left<2I^T_{ij}I^{T\;ij} - 4n^j n^k I^T_{ji}I^{T\;i}_k + n^i n^j n^k n^k I^T_{ij}I^T_{kl}\right>
\end{equation}
The total gravitational flux emitted by the source is the integral of this over the sphere of radius r.
\begin{equation*}
\int Fr^2 \sin \theta d\theta d\phi = \frac{1}{4}\Omega^6\left<\frac{2}{3}I^T_{ij}I^{T\;ij} + \frac{1}{15}(I^{T\;i}_i I^{T\;k}_k  + 2I^T_{ij}I^{T\;ij})\right>
\end{equation*}
which gives
\begin{equation}
L = \frac{1}{5}\Omega^6 \left<I^T_{ij}I^{T\;ij}\right>
\end{equation}
and for a general time dependence of $T$ (till now, $T$ was assumed to be sinusoidal with frequency $\Omega$),
\begin{equation}\label{eq:totalflux}
L = \frac{1}{5} \left<\ddot{\dot{I}}^T_{ij}\ddot{\dot{I}}^{T\;ij}\right> = \frac{G}{5c^5} \left<\ddot{\dot{I}}^T_{ij}\ddot{\dot{I}}^{T\;ij}\right>
\end{equation}

With all this pre-information, an analysis of a binary system can be made. Massive black-hole binaries are the systems that are ultimately being looked at. And the frame is the centre-of-mass frame. Let $y_1{t}$ and $y_2(t)$ be the two trajectories of the masses $m_1$ and $m_2$, and $\vec{y} = \vec{y_1} - \vec{y_2}$, and $r = |\vec{y}|$. The velocities $v_i(t) = \frac{\displaystyle \vec{dy_i}}{\displaystyle dt}$.

The Newtonian equations of motion give:
\begin{eqnarray*}
\frac{d\vec{v_1}}{dt} = - \frac{Gm_2}{r^3}\vec{y};~\frac{d\vec{v_2}}{dt} = - \frac{Gm_1}{r^3}\vec{y}
\end{eqnarray*}
which gives the relative acceleration as $\frac{\displaystyle d\vec{v}}{\displaystyle dt}~=~-\frac{\displaystyle Gm}{\displaystyle r^3}\vec{y}$
 
A simple way of evolving the phase of the gravitational waves emitted by a black-hole Binary system, is to use the energy balance equation. The loss of the centre-of-mass energy is balanced by the total energy emitted as gravitational flux. In the case of circular orbits, this approach is sufficient, as its only needed to find the decrease of the orbital separation $r$.
\begin{equation*}
\frac{d E}{dt} = -L
\end{equation*}
where, $E = -\frac{\displaystyle Gm_1 m_2}{\displaystyle 2r}$ (hence the evolution of $r$ from this).

The outgoing energy is the $L$ that was defined in Eq. \ref{eq:totalflux}. Here the quadrupole moment is ($\mu = m_1m_2/(m_1 +m_2) \& m = m_1 + m_2$);
\begin{equation}\label{eq:IforMBHB}
I^T_{ij} = \mu (y^iy^j - \frac{1}{3} \delta^{ij} r^2 )
\end{equation}

The third derivative needed to calculate the total flux, is easily given by differentiating the above equation. 
\begin{equation}
\frac{d^3 I^T_{ij}}{dt^3} = -4 \frac{Gm\mu}{r^3} (y^i v^j + y^j v^i)
\end{equation}

Replacing this expression into Eq. \ref{eq:totalflux} leads to the 'Newtonian' flux:
\begin{equation}\label{eq:NewtonianFlux}
L = \frac{32}{5}\frac{G^3 m^3 \mu^2}{c^5 r^4} v^2
\end{equation}

It is better expressed in terms of a orbital frequency parameter, $x = \big(\frac{Gm\Omega}{c^3}\big)^{2/3}$ which is of the order $O(1/c^2)$ in the post-Newtonian expansion. Putting in Kepler's Law $Gm = r^3\Omega^2$ the expression for $x$, and $\eta = m_1m_2/m^2$, gives a succinct definition of the Newtonian flux:
\begin{equation}
L = \frac{32}{5} \frac{c^5}{G} \eta^2 x^5
\end{equation}

This is only the Newtonian expression for the flux. Successive Post-Newtonian approximations provide for more accurate expressions. They are described in the next chapter. The core idea of this whole work, is to try to evolve the Post-Newtonian expression using stochastic search methods. The idea germinates from the juxtapositioning of the facts that the field of detection of gravitational waves is replete with usage of stochastic searches, and this is an extension of it in a different direction.

\chapter{LISA}
\label{ch:LISA}

\section{Introduction}

LISA is made of three drag-free spacecrafts, arranged as the vertices of an equilateral triangle, with the length of each side $5\times 10^6$ km. The center of mass of the triangular arrangement follows the Earth in an orbit around the Sun, 20 degree behind the Earth. The plane of this triangular arrangement makes an angle of 60 degrees with the ecliptic. As each of the individual spacecrafts are on different planes, the whole triangular arrangement rotates about itself once every orbital revolution, i.e. with the period of a year. 

The configuration of LISA is such that the triangular arrangement of detectors can be analyzed as a pair of $orthogonal$ two-arm detectors. The motion of the detectors around the Sun introduces a periodic Doppler shift whose magnitude n phase are dependent on the sky position of the source.

If we assume that there are no fluctuations in the arm length, ignore the signal cancellation due to the LISA transfer functions, assume that we can have the data from all three spacecrafts simultaneously, ignore the problem of pointing ahead,  and the frequency of interest is less than the transfer frequency of the detector, i.e. $f << f_* \sim 10^{-2}~Hz$, we are justified in making use of the results of the Low Frequency Approximation (LFA) \cite{LFA}. Of course, as the wavelength becomes comparable to the arm length, the response of the detectors would involve fluctuations of arm length, pointing ahead and the aforementioned signal cancellation, but for our purposes the LFA would suffice \cite{LFA2}.

\section{Mathematical Model}
In the Low-Frequency Approximation, the strain at the detectors (a combination of polarizations weighted by the beam pattern functions, taking into account the measurements at both detectors) due to an incoming GW with polarization $h_{+, \times} (t)$ is:

\begin{equation}
h(t) = h_+ (\chi (t) )F^+ + h_{\times}(\chi (t) )F^{\times},
\end{equation}

where: $\chi (t) = t - R_{\oplus}\sin{\theta} cos({\alpha(t) - \phi})$.

Here, $R_{\oplus} = 1AU \approx 500secs$ is the radial distance to the detector guiding center, $(\theta, \phi)$ are the angular coordinates of the source in the sky, $\alpha(t) = 2\pi f_m t + \kappa,~~f_m = 1/year$ is the LISA modulation frequency and $\kappa$ gives the initial ecliptic longitude of the guiding center. 

In the LFA, the beam pattern functions are essentially a quadrupole antennae. They are defined as:
	
\begin{eqnarray}
F^{+} (t) &= \frac{1}{2} [cos(2\psi)D^+(t;\theta, \phi, \lambda) - \sin(2\psi)D^{\times}(t;\theta, \phi, \lambda)],\\ \nonumber
F^{\times}(t) &= \frac{1}{2}[\sin(2\psi)D^+(t;\theta,\phi,\lambda) + cos(2\psi)D^{\times}(t;\theta, \phi, \lambda)].
\end{eqnarray}

Here, $\psi$ is the polarization angle of the wave. Formally, if $\hat{L}$ is the direction of the binary's orbital angular momentum, and $\hat{n}$ is the direction from the observer to the source ( $180^{\circ}$ to the direction of GW's propagation), then $\psi$ fixes the orientation of the component of $\hat{L}$ perpendicular to $\hat{n}$. The time dependent quantities $D^{+, \times}$ are given in the LFA by \cite{LFA2} as:

\begin{equation}\begin{split}
D^+ (t) =& \frac{\sqrt{3}}{64}\Big[-36\sin^2( \theta)) \sin(2\alpha (t) - 2 \lambda) + (3 + \cos(2\theta)) \\
\big(\cos(2\phi)\lbrace9\sin(2\lambda) -& \sin(4\alpha(t) - 2\lambda)\rbrace +\quad + (\sin(2\phi)\lbrace \cos(4\alpha(t) - 2\lambda) - 9\cos2\lambda))\rbrace\big) \\ & - 4\sqrt{6}\sin(2\theta)\big(\sin(3\alpha(t) - 2\lambda - \phi) - 3\sin(\alpha(t) - 2\lambda + \phi)\big)\Big]
\end{split}\end{equation}

\begin{equation*}\begin{split}
D^\times(t)= & \frac{1}{16}\Big[\sqrt{3} \cos (\theta) \big(9\cos(2\lambda - 2\phi) - \cos(4\alpha(t) - 2\lambda - 2\phi)\big) - \\ 
& 6\sin(\theta)\big(\cos(3\alpha(t) -2\lambda-\phi) + 3\cos(\alpha(t)-2\lambda+\phi)\big)\Big]\quad\;
\end{split}\end{equation*}
\linebreak
Here, $\lambda = 0, \pi$ give the orientation of the two detectors. The GW polarizations up to 2-PN order in amplitude corrections is defined by \cite{LFA2}.

\section{Likelihood Estimator}

Suppose we could create a filter which would output the autocorrelation function of the input. For such a filter, the tranfer function\cite{FFT} would have to be the complex conjugate of the input (in the frequency domain). However, as it turns out to be noncausal, its impossible to make a real filter like this. However, a filter with some frequency delay can be constructed, and would be causal too. This is called the \textit{matched} filter\cite{SigAnalysis1}\cite{SigAnalysis4}. Its so named, because the filter is said to be \textit{matched} with the input as it outputs input's autocorrelation function (with some delay, of course). While \textit{matching} a dummy signal evolved from theoretical templates agains the actual received signal, this is the kind of filtering that is the first idea that comes to one's mind. Towards the construction of such a mathematical tool that can be used for this purpose, first the Natural Scalar product is defined as follows. This, mentionably, involves division of the product of the signal and the template by the spectral density of the noise (all in fourier domain). This indicates, that for the frequencies the noise is high, the 'matching' or vice-versa of the signal and the template should be given less weightage, and vice versa.

\begin{equation}
 \langle h|s\rangle = 2 \displaystyle \int_0^{\infty} \frac{df}{S_n (f)} \big[\tilde{h}(f)\tilde{s}^{*}(f) + \tilde{h}^{*} (f) \tilde{s}(f) \big].
\end{equation}

with vector norm $|h| = {\langle h|h\rangle}^{1/2}$.

\begin{equation}
 \text{Here, } \tilde{h}(f) = \int_{-\infty}^{\infty}h(t)e^{2\pi \iota f t} dt
\end{equation}

is the Fourier Transform of the time domain waveform $h(t)$.  $S_n(f)$ is the one-sided noise spectral density of the detector, and will be defined 
in the next subsection.

Let the signal, in each detector, be $s_i (t) = h_i (t) + n_i (t), i \text{is the index of the detector}$. Assume that the noise $n_i(t)$ is stationary, Gaussian, uncorrelated in each detector and its spectral density is given by $S_n(f)$. Then, we can define the SNR to be:

\begin{equation}
 (SNR)_i = \frac{\langle h|s_i\rangle}{|h|}. \text{i is the index of the detector}
\end{equation}

Given a signal $s(t)$, the likelihood that the true PN coefficients are given by $\overrightarrow{p_n}$ is given by

\begin{equation}\label{eq:Likelihood}
 L(\overrightarrow{p_n}) = C e^{-\langle s-h(\overrightarrow{p_n})|s-h(\overrightarrow{p_n})\rangle /2}
\end{equation}

where $C$ is a normalization constant. From which also follows the reduced log-Likelihood, given as:
\begin{equation}\label{eq:logL}
 \ln L (\overrightarrow{p_n}) = \langle s|h(\overrightarrow{p_n})\rangle - \frac{1}{2} \langle h(\overrightarrow{p_n})|h(\overrightarrow{p_n})\rangle
\end{equation}

Log-Likelihood is going to play the central role in the jump - evaluation of the MCMC. There, the aim would be to maximize in the loglikeliood space \ref{eq:Likelihood}. In the high SNR limit, the error in the determination would be $1\sigma^2$. The SNR and the log-Likelihood are related by:
\begin{equation}
lnL \approx SNR^2 / 2
\end{equation}

\section{Detector Noise}
The Noise Spectral Density, has two chief components. The instrumental noise, and the confusion noise. To model the instrumental noise, the expression for the standard one-sided noise spectral density for the LISA is used\cite{instr_noise_LISA}:
\begin{equation}
S^{ins}_n(f) = \frac{1}{4L^2} \big[S^{pos}_n(f) + 2\big(1 + \cos^2 \big(\frac{f}{f_*}\big) \big) \frac{S_n^{acc}(f)}{(2\pi f)^4} \big]
\end{equation}
where, $L = 5 \times 10^6 km$ is the arm length of LISA, $S^{pos}_n(f) = 4 \times 10^{-22} m^2 /Hz$ and $S_n^{acc}(f) = 9 \times 10^{-30} m^2/s^4/Hz$ are the position and acceleration noise respectively.

The quantity $f_* = 1 / (2\pi L)$ is the mean transfer frequency for the LISA arm. To the above formula, a random gaussian is given as input, to generate the simulated instrumental - noise spectral density.

\begin{figure}[tp]
\centering
\includegraphics[keepaspectratio=true,width=4.5in]{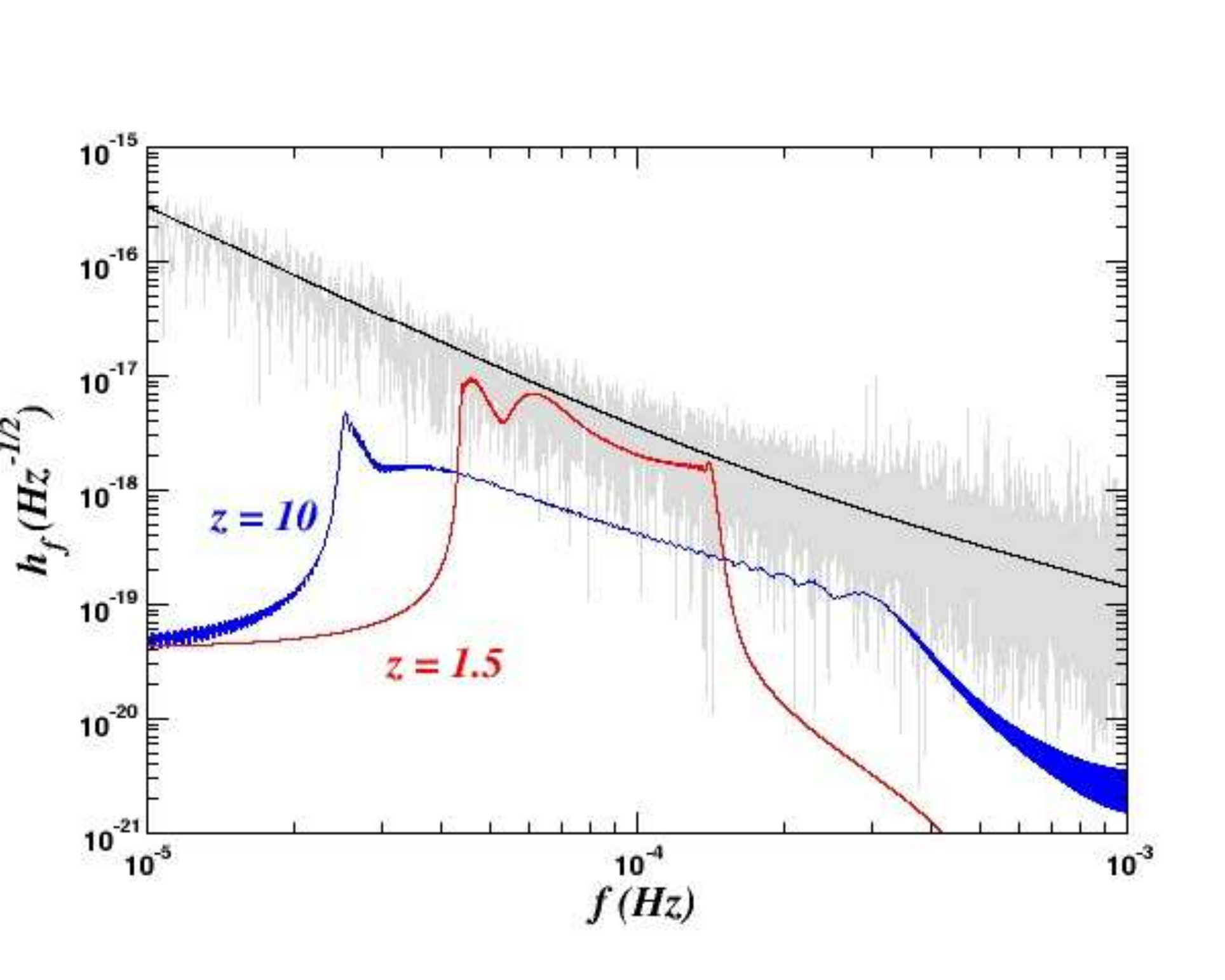}
\caption{LISA's noise curve.}
\label{fig:LISA_noise}
\end{figure}

The plenitude of unresolvable galactic binaries which constitute the galactic foreground produce a noise source which is called the confusion noise. This noise is can not be assumed to be stationary and Gaussian. The binaries which are bright enough to be individually resolved, are resolved and removed beforehand. To model the remaining galactic noise, Nelemans, Yungelson and Zwart (NYZ) came up with the following expression\cite{galactic_noise1}\cite{galactic_noise2}:
\begin{eqnarray}
S_n^{conf}(f) &=& 10^{-44.62} f^{-2.3}, 10^{-4} < f \leq 10^{-3} \nonumber \\
              &=& 10^{-50.92} f^{-4.4}, 10^{-3} < f \leq 10^{-2.7} \nonumber \\
              &=& 10^{-62.8} f^{-8.8}, 10^{-2.7} < f \leq 10^{-2.4} \\
              &=& 10^{-89.68} f^{-20}, 10^{-2.4} < f \leq 10^{-2} \nonumber \\ 
\end{eqnarray}
And so the total spectral density of noise is given by: $S_n(f) = S_n^{ins}(f) + S^{conf}_n(f)$. A plot of the noise is given in the Fig. \ref{fig:LISA_noise}

\section{The Gravitational Waveform}

The two harmonics of the wave can be written as:

\begin{equation}
h_{+,\times}(t) = Re { \displaystyle\sum^{k} H^{(k)}_{+,\times}(t)e^{\imath k \Phi_{orb}}},
\end{equation}

where, 'k' denotes the order of the harmonic, $\phi_{orb}$ is the orbital phase and $H^{(k)}_{+,\times}$ are the amplitude components associated with each harmonic. The strongest harmonic of the wave is the one corresponding to $k = 2$, given by the quadrupole moment of the source. So, in the restricted post-Newtonian approximation, we ignore all terms other than the quadrupole term for the amplitude, and expand the phase corrections of the wave up to the order of $~(v/c)^{2n}$, ie n-PN order. Also, the approximation neglects the phase modulation brought in by the other order amplitude corrections. In this approximation, the GW polarizations are given by \cite{higher_harmonics}:

\begin{equation}\label{eq:harmonics}
h_{+, \times} = \frac{2Gm\eta}{c^2 D_L}x[H^{(0)}_{+,\times} + x^{1/2}H^{(1/2)}_{+,\times} + xH^{(1)}_{+,\times} + x^{3/2}H^{(3/2)}_{+,\times} + x^2 H^{(2)}_{+,\times}].
\end{equation}

Here, $m = m_1 + m_2$, the total mass of the binary, $\eta = m_1 m_2 / m^2$ is the reduced mass ratio, and $D_L$ is the luminosity distance of the source. The post-Newtonian parameter $x = (Gm\omega / c^3)^{2/3}$, where $\omega = d{\Phi_{orb}}/dt$.

In Eqn \ref{eq:harmonics}, the $H^{(n)}$ include the post Newtonian corrections to the amplitude and the extra phase harmonics. But for our purpose, it suffices to work in the restricted PN approximation, which means including only $H^{0}_{+,\times}$ terms. So, the GW polarizations are given by \cite{Hpol}:

\begin{eqnarray}
h_+ &=& \frac{2Gm\eta}{c^2 D_L} (1+cos^2(\iota))x cos(\phi),\\ \nonumber
h_{\times} &=& -\frac{4Gm\eta}{c^2 D_L}cos(\iota) x \sin(\phi).
\end{eqnarray}

The inclination of the orbit of the binary is defined as $cos(\iota) = \hat{L} \cdot \hat{n}$.

In the adiabatic approximation, the evolution of the phase of the wave, and the instantaneous velocity is governed by

\begin{eqnarray}
\frac{dv}{dt} &=& - \frac{F(v)}{m E'(v)}.\\
\frac{d\phi}{dt} &=& \frac{2 v^3}{m} \nonumber
\end{eqnarray}
 
Where, $v = (\pi m f)^{1/3}$ is the instantaneous velocity, $E'(v) = dE/dv$ is the derivative of the orbital enerdy with respect to the velocity. $F(v)$ is the gravitational wave flux function. For a test-mass particle in circular orbit around a Schwarzschild black hole, and exact expression for the orbital energy is given in \cite{E}, the derivative of which is given as:

\begin{equation}
 E'(v) = -\eta v \frac{1-6 v^2}{(1-3 v^2)^{3/2}}
\end{equation}

We can see that this equation gives an Energy extremum at $v = 1/\sqrt{6}$, which is also the velocity at the last stable orbit $v_{lso}$.
For the gravitational Flux function, whose Newtonian part was derived earlier in the Section \ref{sec:NewtonianFlux}, there is a 5.5PN expression\cite{PN5.5-1} \cite{PN5.5-2} \cite{PN5.5-3} \cite{PN5.5-4} \cite{PN5.5-5} \cite{PN5.5-6}:

\begin{equation}\label{eq:g_flux}
F_{T_n}(v) = F_N (v) \Big[ \sum_{k=0}^{11} a_k v^k + ln (v) \sum_{k=6}^{11} b_k v^k + {O} (v^{12}) \Big].
\end{equation}

\begin{equation}
 F_N (v) = \frac{32}{5} \eta^2 v^{10}.
\end{equation}

\begin{equation*}\begin{split}
a_0 &= 1,\quad a_1 = 1,\quad a_2 = -\frac{\displaystyle 1247}{\displaystyle336}, \quad a_3 = 4\pi, \quad a_4 = -\frac{\displaystyle 44711}{\displaystyle 9072}, \quad a_5 = -\frac{\displaystyle 8191\pi}{\displaystyle 672}, \\
a_6 &= \frac{\displaystyle 6643739519}{\displaystyle 69854400} - \frac{\displaystyle 1712\gamma}{\displaystyle 105} + \frac{\displaystyle 16\pi^2}{\displaystyle 3} - \frac{\displaystyle 3424ln(2)}{\displaystyle 105}, \quad a_7 = -\frac{\displaystyle 16285\pi}{\displaystyle 504},\\
a_8 &= -\frac{\displaystyle 323105549467}{\displaystyle 3178375200} + \frac{\displaystyle 232597\gamma}{\displaystyle 4410} - \frac{\displaystyle 1369\pi^2}{\displaystyle 126} + \frac{\displaystyle 39931 ln(2)}{\displaystyle 294} - \frac{\displaystyle 47385 ln(3)}{\displaystyle 1568},\\
a_9 &= \frac{\displaystyle 265978667519\pi}{\displaystyle 745113600} - \frac{\displaystyle 6848\gamma\pi}{\displaystyle 105} - \frac{\displaystyle 13696\pi ln(2)}{\displaystyle 105},\\
a_{10} &= -\frac{\displaystyle 3500861660823683}{\displaystyle 2831932303200} + \frac{\displaystyle 916628467\gamma}{\displaystyle 7858620} - \frac{\displaystyle 424223\pi^2}{\displaystyle 6804} - \frac{\displaystyle 83217611ln(2)}{\displaystyle 1122660} + \frac{\displaystyle 47385ln(3)}{\displaystyle 196},\\
a_{11} &= \frac{\displaystyle 8399309750401\pi}{\displaystyle 101708006400} + \frac{\displaystyle 177293\gamma\pi}{\displaystyle 1176} + \frac{\displaystyle 8521283\pi ln(2)}{\displaystyle 17640} - \frac{\displaystyle 142155\pi ln(3)}{\displaystyle 784},\\
\end{split}\end{equation*}

 and

\begin{equation*}
 b_6=-\frac{\displaystyle 1712}{\displaystyle 105}, b_7=0, b_8 = \frac{\displaystyle 232597}{\displaystyle 4410}, b_9 = -\frac{\displaystyle 6848\pi}{\displaystyle 105}, b_{10} = \frac{\displaystyle 916628467}{v\displaystyle 7858620}, b_{11} = \frac{\displaystyle 177293\pi}{\displaystyle 1176}.\\
\end{equation*}

\chapter{Markov Chain Monte Carlo}

\section{Introduction}
Stochastic search algorithms inspired by physical and biological systems are applied to the problem of optimization in multiple dimensions, with multiple local optima. For this type of systems, greedy deterministic search algorithms tend to halt at local optimum, usually requiring random restarts to obtain solutions of acceptable quality. Stochastic search for constrained optimization can be viewed as a unifying paradigm for expressing the fundamental laws governing the behavior of natural systems. Physical systems can be modeled as “detecting” local potential fields and “seeking” states of low free energy. Quantum systems can be modeled as choosing stochastically among local extrema of the action integral \cite{Shankar}. 

The current problem at hand is essentially of maximizing a function (the log-Likelihood), in the Post Newtonian coefficient space. The non-stochastic search for the coefficients with the aim of maximizing over the log-Likelihood function, can prove to be extremely computationally intensive for this case. In the non-stochastic approach, the whole space is first divided into a grid, and then within each grid independent searches proceed. For example, if searching in two dimensions, the surface would be divided into a hexagonal grid (closest packing):
\begin{figure}[h]
\centering
\includegraphics[keepaspectratio=true,width = 3.5in]{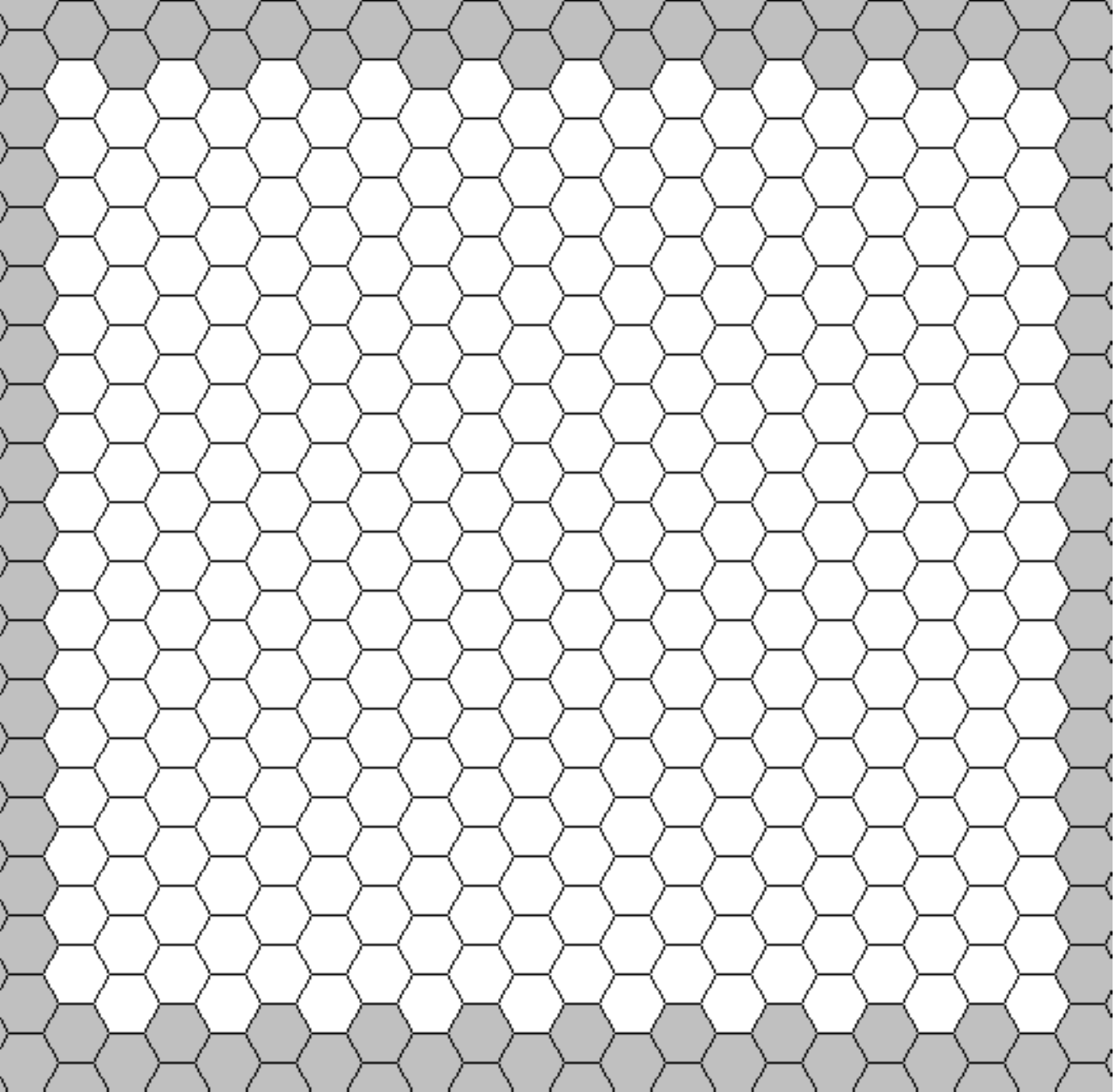}
\caption{hexagonal grid for 2-D non-stochastic search}
\label{fig:hexgrid}
\end{figure}

In three dimensions, a FCC lattice is used, as it has the highest packing fraction. But, as we proceed to higher dimensions, the most efficient packing formations itself are not known - which is the basic information on the basis of which the space is covered by a grid. In higher dimensions, due to the size of the search space, the non-stochastic algorithm would definitely not be computationally very feasible.
Hence, the stochastic search methods were explored for this particular problem.

Natural systems provide a rich source of analogies for constructing efficient approaches to complex search, optimization, and learning problems. Any problem of free energy minimization can be recast as an optimization or statistical inference problem. The energy and energy states of the system are interpreted as the objective function to be minimized or the probability distribution to be simulated. 

A fairly popular approach applied to derive an approximate solution to the target problem is the Markov Chain Monte Carlo, or MCMC. For instance, applying MCMC to a mechanical system, a stochastic simulation of the system is constructed in which the long run frequency with which each solution is visited is given by the distribution that minimizes free energy, ie the Boltzmann distribution. Because low energy states have been defined as “good” according to the target objective function or probable according to the target distribution, the simulated system evolves over time to spend more of its time at good solutions of the target problem. 

There exists a considerable literature documenting successful application of MCMC methods to a wide variety of problems [e.g., \cite{egMCMC}]. A variety of MCMC samplers have been constructed for any given problem by varying the sampling distribution subject to the local reversibility conditions that ensure convergence to the optimal distribution distribution. 

Although the long-run frequency distribution is identical for any MCMC sampler satisfying the ergodicity conditions, different samplers on the same surface can vary widely in their dynamic behavior. Especially the speed with which a sampler reaches the optima and the ease with which it escapes local hills or valleys. Randomized restarts help to avoid the tendency to become stuck on a local peak, but lack of mobility remains a problem for highly complex and multimodal surfaces. 

Various approaches have been proposed to improve performance of MCMC samplers. If global information is available about the surface, it can be used to inform sampling and thus increase efficiency. For example, samplers can be made by hybridising Differential Evolution and Markov Chain Monte Carlo \cite{DE_storn_price} \cite{DE_braak}, using a population of MCMC samplers to assess the variability in results from different runs of the sampler. It has been suggested that multiple parallel runs might provide an opportunity to improve performance by exchanging information among solutions \cite{DE_braak}. However, due to computational intensity, these approaches were found impractical for the task at hand.

\section{A Little background}

Let us imagine that we need to calculate the expected value of a function $\phi (x)$, where $x$ is a random vector of dimensionality $k$, with a probability density $f(x)$. Then, this expected value is given as (taking the reasonable assumption $E[|\phi(x)|] < \infty$ ):
\begin{equation}
E[\phi(x)] = \int_{\Re^k} \phi(x) f(x) dx
\end{equation}

In the case where analytical integration is not feasible, and numerical integration is tedious, the \textit{Monte Carlo} method is employed. It involves generating a sequence of random vectors ${x_n}$ with the same distribution $f(x)$, and the strong law of large numbers gives:
\begin{equation}
E[\phi(x)] = \lim_{n\rightarrow \infty} \frac{1}{n} \sum^n_{i=1} \phi(x_i)
\end{equation}
as the estimate for the expected value. 

But what if sampling from $f(x)$ is practically difficult or infeasible. At this point the \textit{Markov Chain Monte Carlo} comes into the picture. For simplicity, consider a system with a finite number of possible states, $x_1, x_2, x_3, \dots x_n$. To each state, assign a probability, $p_i = f(x_i)$.

Suppose we have to calculate the expected value of $\phi(x)$, as above. Therefore, 
\begin{equation}
E[\phi] = \sum^n_{i=1} \phi(x_i) f(x_i)
\end{equation}

The probability of transition to a particular state, depends only on the immediately preceding state. This identifies the chain as Markovian. Formally, 
\begin{equation}
P(x(t) = x_j|x(t-1)=x_{j1}, x(t-2)=x_{j2},\dots ) = P(x(t) = x_j|x(t-1) = x_i) = p_{ij}
\end{equation}

The matrix $P$ with elements $p_{ij}$ is called the Markov matrix.

A state j is said to be accessible from a different state i ($i \rightarrow j$) if, given that we are in state i, there is a non-zero probability that at some time in the future, the system will be in state j. Formally, state j is accessible from state i if there exists an integer $n\geq0$ such that
\begin{equation}
    P(x_{n}=j | x_0=i) > 0
\end{equation}
A state i is said to communicate with state j if it is true that both i is accessible from j and that j is accessible from i. A set of states C is a communicating class if every pair of states in C communicates with each other, and no state in C communicates with any state not in C. A Markov chain is said to be \textit{irreducible} if its state space is a communicating class; this means that, in an irreducible Markov chain, it is possible to get to any state from any state.

Finally, a markov chain is ergodic in a State Space if all of the space is irreducible with respect to the chain. And there exist numbers $\phi_i \geq 0, \sum_i \pi_i = 1, and~\pi_j = \sum_i \phi_i p_{ij}$.

\section{Simulated Annealing \cite{Simulated_Annealing1} \cite{Simulated_Annealing2}} 
Simulated annealing (SA) is a generic probabilistic meta-algorithm for the global optimization problem, namely locating a good approximation to the global maximum of a given function in a large search space. It is often used when the search space is discrete provided that the goal is to find an acceptably good solution in a fixed amount of time, rather than the best possible solution.

The name and inspiration come from annealing in metallurgy, a technique that involves controlled cooling of a pre-heated material to increase the size of its crystals and reduce their defects. The heat causes the atoms to become unstuck from their initial positions (a local minimum of the internal energy) and wander randomly through states of higher energy; the slow cooling gives them more chances of finding configurations with lower internal energy than the initial one. The distribution of states, as a function of Energy, is given by the Boltzmann distribution:
\begin{equation}\label{eq:boltzmann}
f_E = 2 \big( \frac{E}{\pi(kT)^3}\big) e^{\frac{\displaystyle -E}{\displaystyle kT}}
\end{equation}
Obviously, as the T increases, the probability of more particles being at a higher value is E increases. That means, that the particles explore the whole energy band completely (in other words, are distributed in a wide region of the energy band). Now, for the stochastic search, the MetroPolis Hastings ratio [Eq. \ref{eq:MetroHastings}] is raised to the power $1/T$. This lowers the peaks on the H surface too. 
\begin{figure}[h]
\centering
\includegraphics[keepaspectratio=true,width = 3.5in]{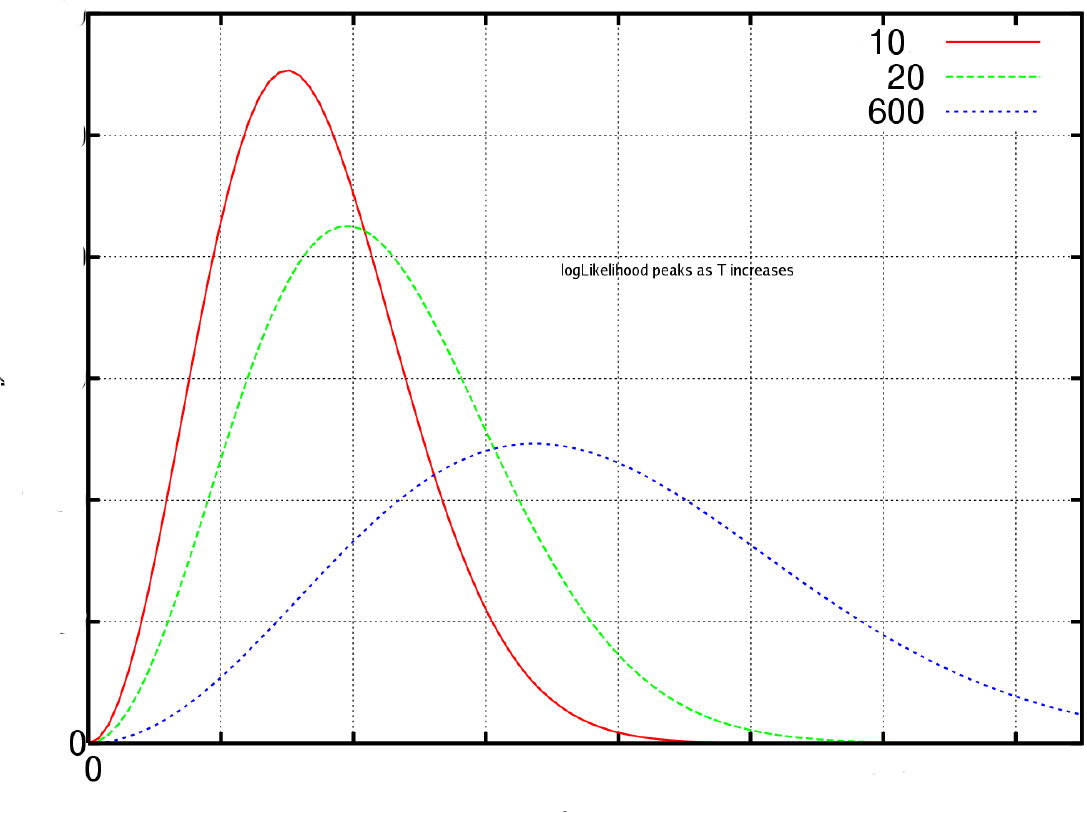}
\caption{the Likelihood peaks, as the temperature increases}
\label{fig:effectofT}
\end{figure}

By analogy with the physical process, each step of the Simulated Annealing algorithm replaces the current solution by a random "nearby" solution, chosen with a probability that depends on the difference between the Likelihood estimater values and on a global parameter T (called the \textit{temperature}), that is gradually decreased during the process. The net affect is that the peaks are lowered and broadened, as shown in the figure above. The dependency is such that the current solution changes almost randomly when T is large, but increasingly "uphill" as T goes to zero. The allowance for "downhill" moves saves the method from becoming stuck at local maxima — which is the prime negative point against \textit{greedy} algorithms. A \textit{cooling schedule} defines how the temperature falls. It has to be adjusted to allow enough time for the initial \textit{burn-in} phase, as the option of De-initializing of the chain \cite{De-initializing} is not availed.

The importance of the cooling schedule can not be over-emphasised. Suppose that the temperature is decreased too rapidly. Then, if the chain gets once stuck at a secondary maximum, it would not even get time to come out of it, and the temperature would have fallen already. That would mean that the probability of the chain getting out of the secondary maximum would fall sharply with time. And, if the temperature falls too slowly, then the chain has just too much time at each peak, including the global maxima, and it might just wander off.
Thus, the way the fall in the temperature is regulated changes a lot in the destiny of the chain.

The method that is used here, is an adaptation of the Metropolis-Hastings algorithm\cite{Metro_Hastings}, a Monte Carlo method to generate sample states of a thermodynamic system, invented Metropolis et al. 1953.

\subsection{Thermostated Annealing\cite{Search_MBHB_LISA}}

A slight modification of Simulated Annealing works better, in case where high Temperature values are needed to keep the chain from getting stuck, at all times. This goes by the indicative name of $thermostated$ annealing.

When the temperature falls below a certain predecided threshold, the Temperature of the system is $increased$ and levelled off. The threshold is fixed in terms of the highest SNR attained till that point. As the logLikelihood \ref{eq:logL} is related to the Signal to Noise ratio, by $\ln L \approx \frac{1}{2}SNR^2$, this relation readily translates into one in terms of the log-Likelihood. This injection of heat ensures that the highest possible temperature is used, which would help the chain to move out of local maxima. As raising to higher temperature would cause lowering of the peaks, the movement about the surface would be consequently easier.

\section{The Implementation}

Let $A_m$ denote the set of the coefficients of $v^i$, and $B_n$ denote the set of the coefficients of $\ln (v)v^i$ in the 5.5PN expansion of the gravitational Flux function\ref{eq:g_flux} that are included in the search space (the indexing is done backwards). This means, that if $m = 2, n = 1$, then last $m$(=2) a's\ref{eq:g_flux} and last $n$(=1) b's\ref{eq:g_flux} are included in the search space.

The values of $m$ and $n$ are changed, and the results were observed. The underlying implementation of MCMC remained more or less the same. So, the search is always in a $m+n$ dimensional space.
\subsection{The Algorithm}
The algorithm proceeded as follows:

Let us define a vector 
\begin{equation}\label{eq:defmn}
\vec{x} = {a_{12 - i}, b_{12 - j}; 1\leq i \leq m, 1\leq j \leq n} 
\end{equation}

\textit{Step 1:} An arbitrary starting point is chosen for $\vec{x}$.

\textit{..iteration begins..}

\textit{Step 2:} A draw is made for the displacement in a single dimension of the search space, from a zero - mean Gaussian proposal distribution. (As all the coefficients in the PN expansion are non-correlated, we can take jumps in one dimension at a time, cyclically for all 'elements of' x). The new point, generated by adding the jump to $x_i$, where $i$ changes cyclically over $[1,m+n]$ with each iteration.
i.e. $\vec{y} = \{ x_1, \dots , x_i + draw, \dots\} $

\textit{Step 3:} The Metropolis - Hastings ratio\cite{Metro_Hastings}: 
\begin{equation}\label{eq:MetroHastings}
H = \big[ \frac{\displaystyle\pi(\vec{y})p(s|\vec{y})q(\vec{x}|\vec{y})}{\displaystyle(\vec{x})p(s|\vec{x})q(\vec{y}|\vec{x})} \big]^{1/T}
\end{equation}
is evaluated. Here, $\pi(\vec{x})$ are the priors of the coefficients, $p(s|\vec{x})$ is the likelihood of $\vec{x}$ denoting the true value of coefficients, given by comparing the signal evolved using $\vec{x}$, to the signal given by the known 5.5PN approximation\ref{eq:Likelihood}. And the last term, $q(\vec{y}|\vec{x})$ is the proposal distribution.

As a symmetric sampling distribution is used, $q(\vec{x}|\vec{y})~=~q(\vec{y}|\vec{x})$. 

\textit{Step 4:} The $jump$ is taken, with a probability of $min(1, H)$.

\textit{Step 5:} The temperature is configured for the next iteration. The cooling schedule that is chosen works as follows:

Lets say that the chain runs $N$ times, and initial temperature be $T_0$. Then $nheat (< N)$ is chosen, which would determine the number of steps for which the surface would be heated. Let $k$ denote the index of iteration, then the temperature is decided by:
$\\ \text{if}(k < nheat)\\
\text{\qquad} \{ \\
\text{\qquad} heat = \displaystyle T_0^{(1-\frac{k}{nheat})} \\
\text{\qquad} \text{if}(heat \leq T_{thresh}) heat = T_{thresh} \\
\text{\qquad} T_{last} = heat \\
\text{\qquad}\} \\
\text{\qquad} \text{else~if}(nheat < k < N) \\
\text{\qquad \qquad} \{ \\
\text{\qquad \qquad} heat = T_{last}^{(1 - \frac{k}{N-nheat})}\\
\text{\qquad \qquad} \} \\
\text{\qquad \qquad} \text{else}\\
\text{\qquad \qquad \qquad} heat = 1.0\\
\text{and}, \\
T_{thresh} = SNR^2 / 100\\$

Then, back to \textit{Step 2.}

\subsection{The Approach}\label{subsec:approach}
 
The first step was to start with a search for $m = 1, n = 0$, keeping all the other coefficients fixed at their original values. The true signal was taken to be the 5.5PN one. Then the dimensionality of the search was increased in steps of 1, alternately increasing $m$ and $n$ by 1.

This was done till the Chains were able to successfully converge at the true values. Then this is repeated, including noise in the original signal $\vec{s}$. It is important to understand the role played here by the cooling schedule here. The surface plot / contour plots for the surface were generated, to help with the cooling schedule. Initially a simple schedule was introduced, where the temperature fell exponentialy to 1, over the whole length of the chain. Then another schedule was implemented, in which the temperature fell exponentially to 1 by 15000 steps, and the last 5000 steps were at a constant temperature of 1, and this was done for the chain to search locally (around the point it would be at at the end of the 15000 steps), and yield more accurate results. Also, in this step the number of steps in the chain was increased. Then a schedule was implemented such that the temperature exponentially fell, but never below a particular threshold. This was done after inspecting the contour plots, which showed a row of maxima and it was expected that the chain would get stuck in them. So, a higher temperature would be required always for it to be freed.

The important fact about the sampling scheme is, that it produces a Markov chain with a stationary distribution equal to the posterior distribution of interest, $p(\vec{x}|s)$, regardless of the choice of proposal distribution\cite{egMCMC}, although a poor choice of the proposal distribution would result in the algorithm taking a very long time to converge to the stationary distribution (known as the $burn-in$ time).
The jumps are scaled\cite{scaling} by the square-root of the heat. Since the chain is not de-initialized \cite{De-initializing}, elements of the Markov chain produced during the $burn - in$ phase have to be discarded as they do not represent the stationary distribution. Also, as the dimensionality of the search space increases, this $burn - in$ time can be very long.

If the search parameters are correlated, the chain is not very efficient in exploring the whole search space, but in this case as the coefficients are all independent, the chain was expected to have explored the search space thoroughly.
 
%


\chapter{Results}

The sources which are expected to be observed can be typically divided in to two distinct groups. First, where we see coalescence and second, where we dont. For the purpose here, either of them would do, as the more engaging matter here is about the accuracy of the PN-template available. So, a system where coalescence is not observed is chosen. A list of the parameters of the system that was chosen follows:
\linebreak

\begin{tabular}{|m{4cm}|m{4cm}|}
\hline Parameters      & Values \\ \hline
$m_1/M_{\cdot}$        & $5 \times 10^6$\\
$m_2/M_{\cdot}$        & $10$ \\
$\theta / rad$         & $0.6842$ \\
$\phi / rad $          & $ 2.5791$ \\
$ t_c /yr$             & $0.458333$ \\
$\iota / rad$          & $0.9273$ \\
$\psi / rad$           & $1.4392$ \\
$z$                    & $10$ \\
$D_L / Gpc$            & $10^{-1}$\\
$\varphi_c /rad$       & $0.3291$ \\
$SNR$                  & $\sim 40$\\ \hline
\end{tabular}

\begin{figure}[h]
\centering
\includegraphics[keepaspectratio=true,width = 5.5in]{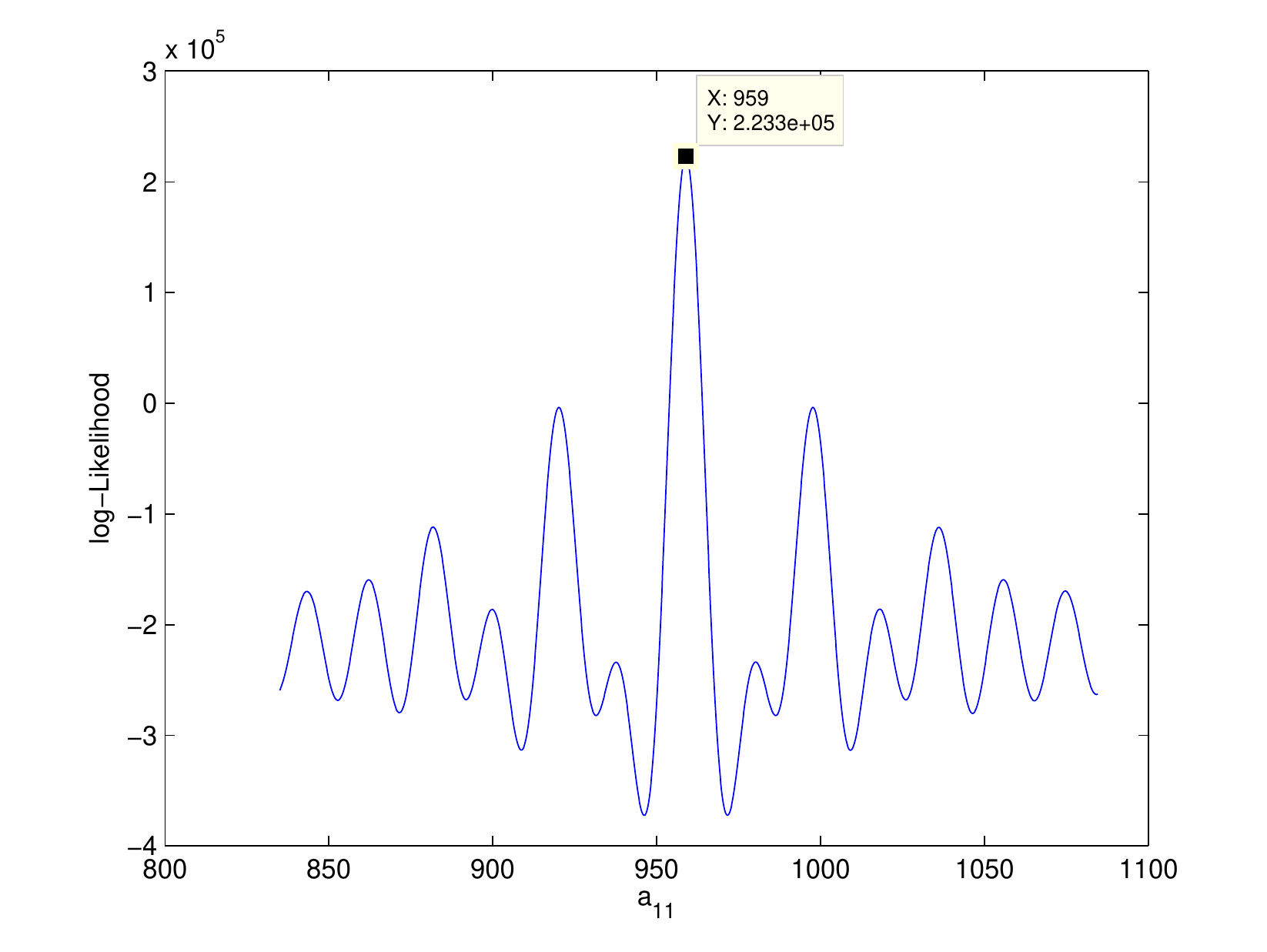}
\caption{Loglikelihood v/s coefficient $a_{11}$.}
\label{fig:a11}
\end{figure}

\begin{figure}[h]
\centering
\includegraphics[keepaspectratio=true,width = 5.5in]{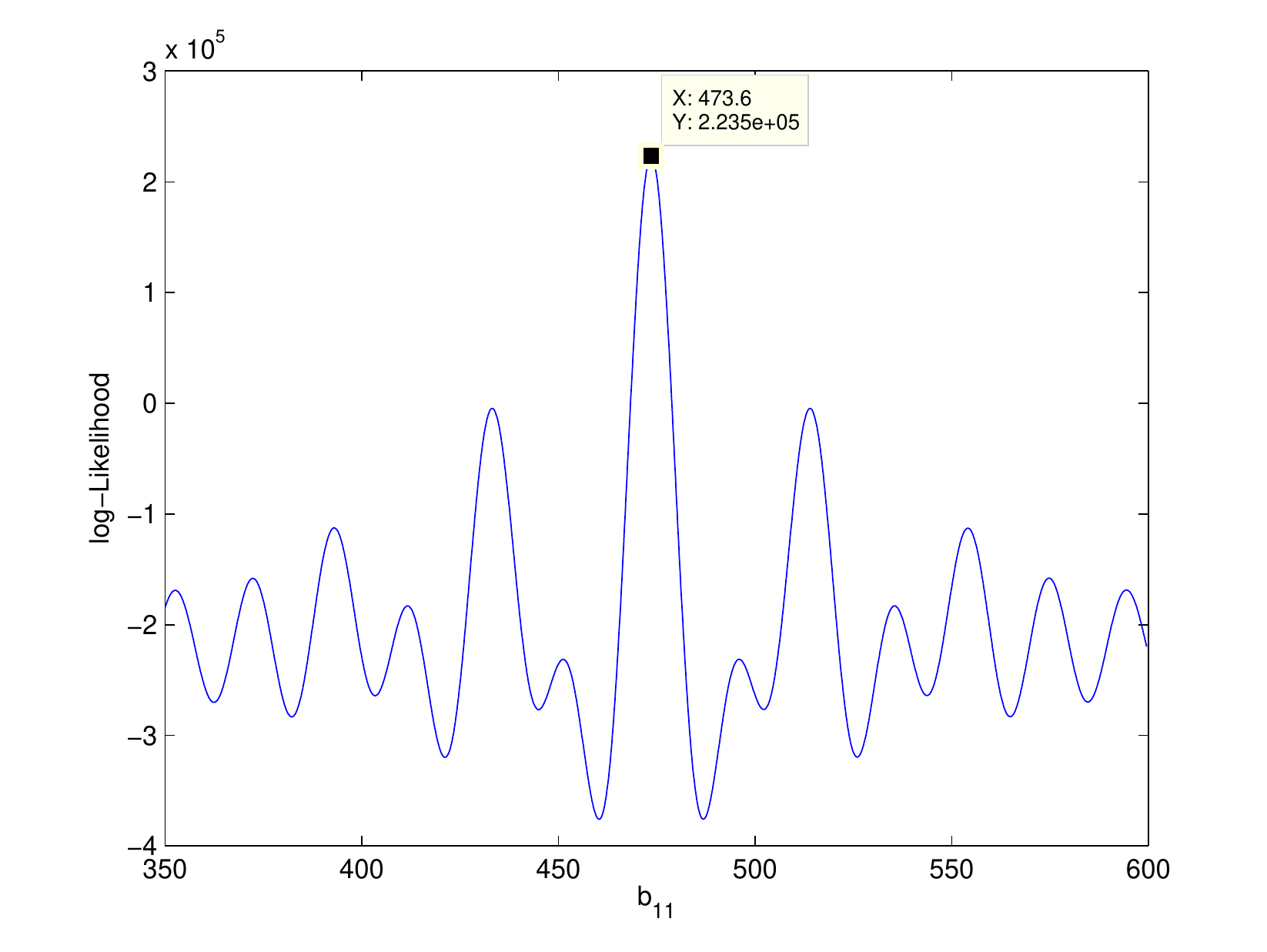}
\caption{Loglikelihood v/s coefficient $b_{11}$.}
\label{fig:b11}
\end{figure}

The first step was to start with a search for $a_{11}$, keeping all the other coefficients fixed at their original values. For this search, the one-dimensional surface was plotted, and Figure \ref{fig:a11} was obtained. Also, at the same time, the one-dimensional surface was plotted for $b_{11}$ [Figure {\ref{fig:b11}}]. The purpose of plotting these was to make sure that the starting point of the chain is not chosen too close. Also, to predict the heat required in the beginning could be estimated better with the help of these. It is noteworthy, that the true Signal was taken without the noise. So, this was really a test run per se.

\begin{figure}[tp]
\centering
\includegraphics[keepaspectratio=true, width = 5.5in]{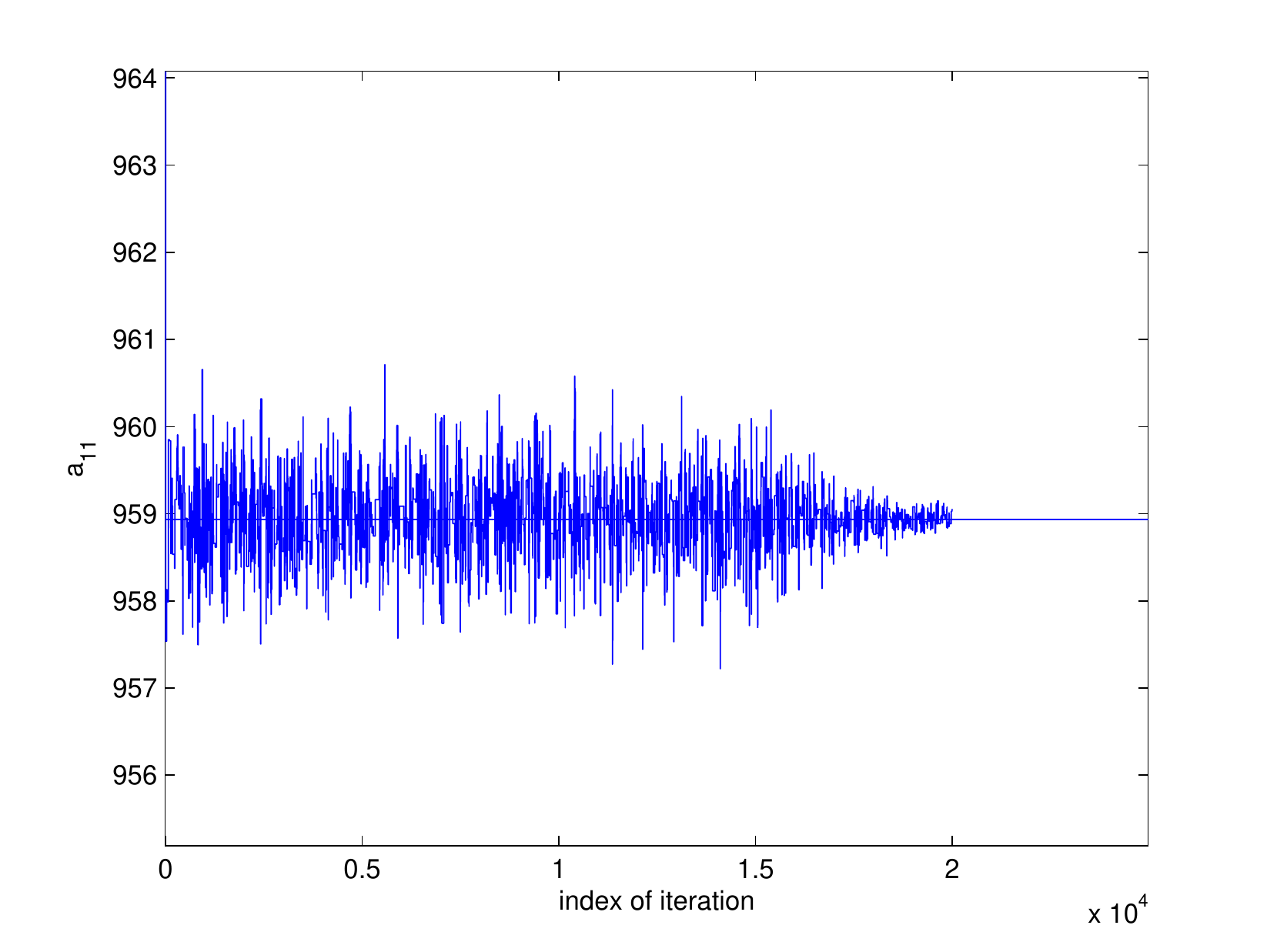}
\caption{Plot of the chain searching for $a_{11}$ in $absence$ of noise.}
\label{fig:chain_a11_wn}

\centering
\includegraphics[keepaspectratio=true, width = 5.5in]{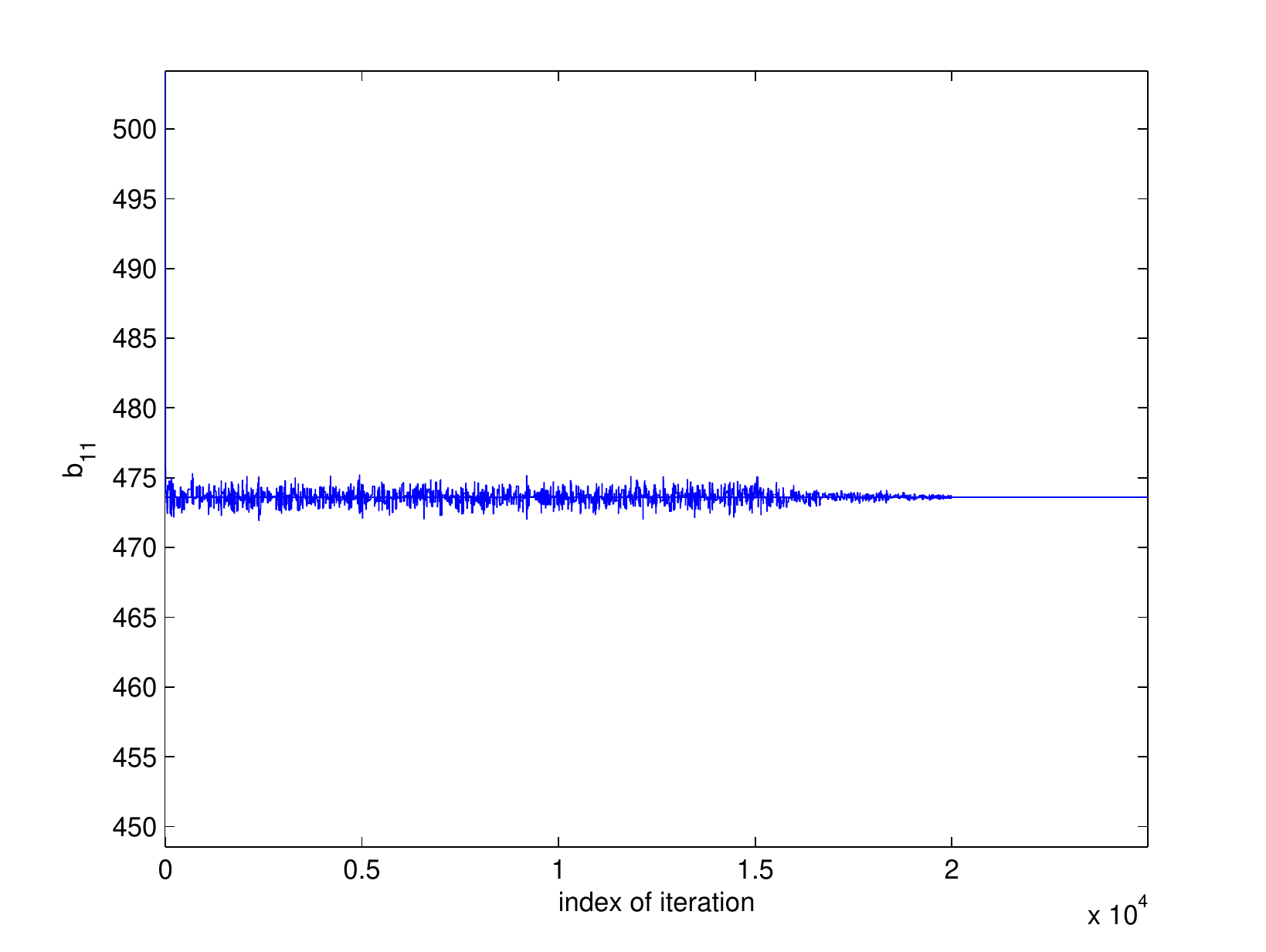}
\caption{Plot of the chain searching for $b_{11}$ in $absence$ of noise.}
\label{fig:chain_b11_wn}
\end{figure}

As these curves are pretty smooth, and not too steep, it was predictable that the Markov Chain should be able to converge on the true value. The SNR ranged around $600$, which is also a pretty high value. Although, of course, the original signal being deviod of noise, the term Signal-to-Noise Ratio does not hold much relevance. The results from the Markov Chains was extremely rapid convergence. This is given in the Fig.\ref{fig:chain_a11_wn} and Fig.\ref{fig:chain_b11_wn}. The constant blue horizontal line denotes the true value (known) of the coefficient. 

As the chains converged at the true value of $a_{11}$ and $b_{11}$, the most obvious next thing to do would be to check if the same result is obtained in presence of noise in the original Signal.
\pagebreak

\begin{figure}[tp]
\centering
\includegraphics[keepaspectratio=true, width = 5.5in]{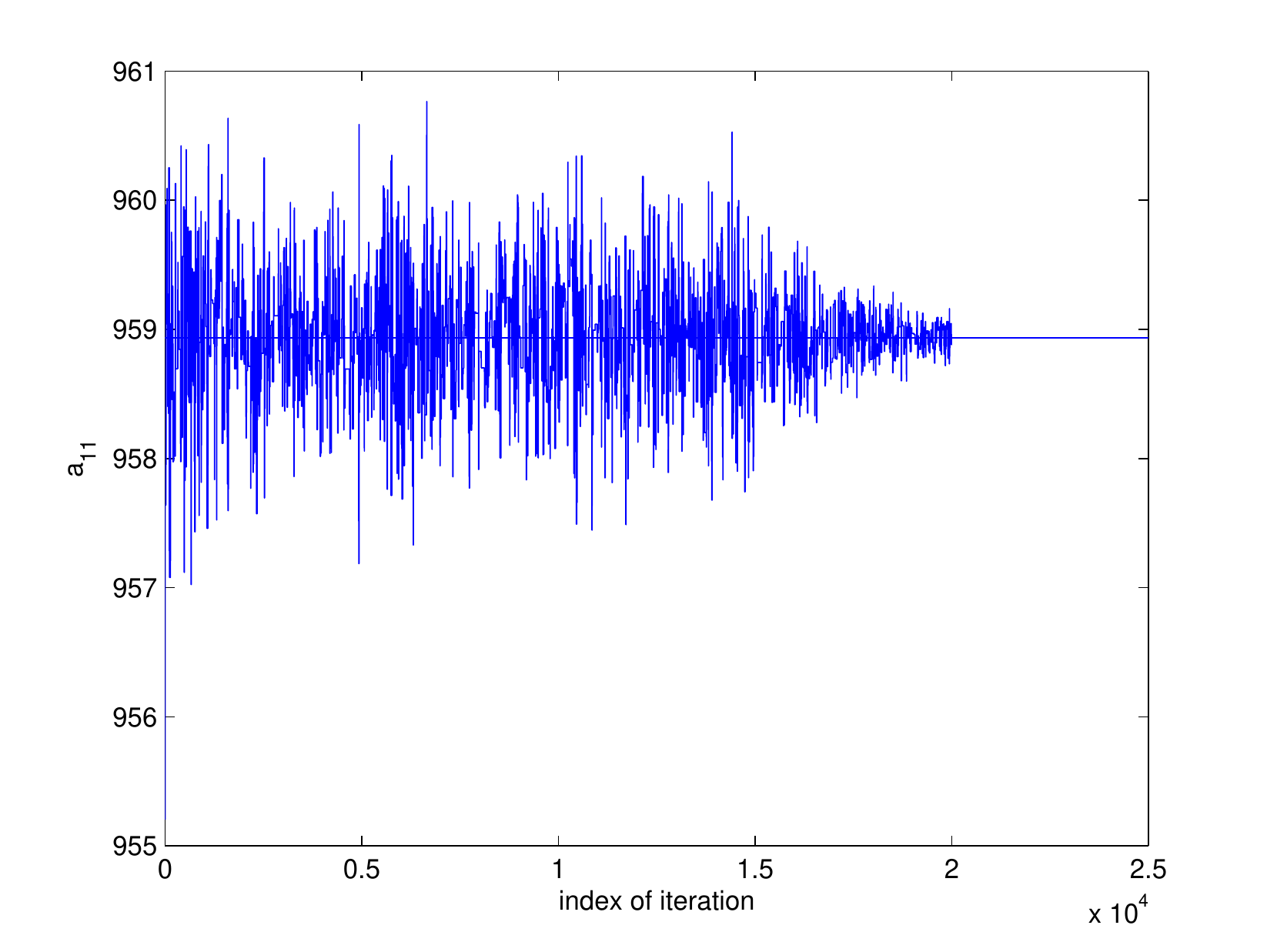}
\caption{Plot of the chain searching for $a_{11}$ in $presence$ of noise.}
\label{fig:chain_a11}
\end{figure}

\begin{figure}[h]
\centering
\includegraphics[keepaspectratio=true, width = 5.5in]{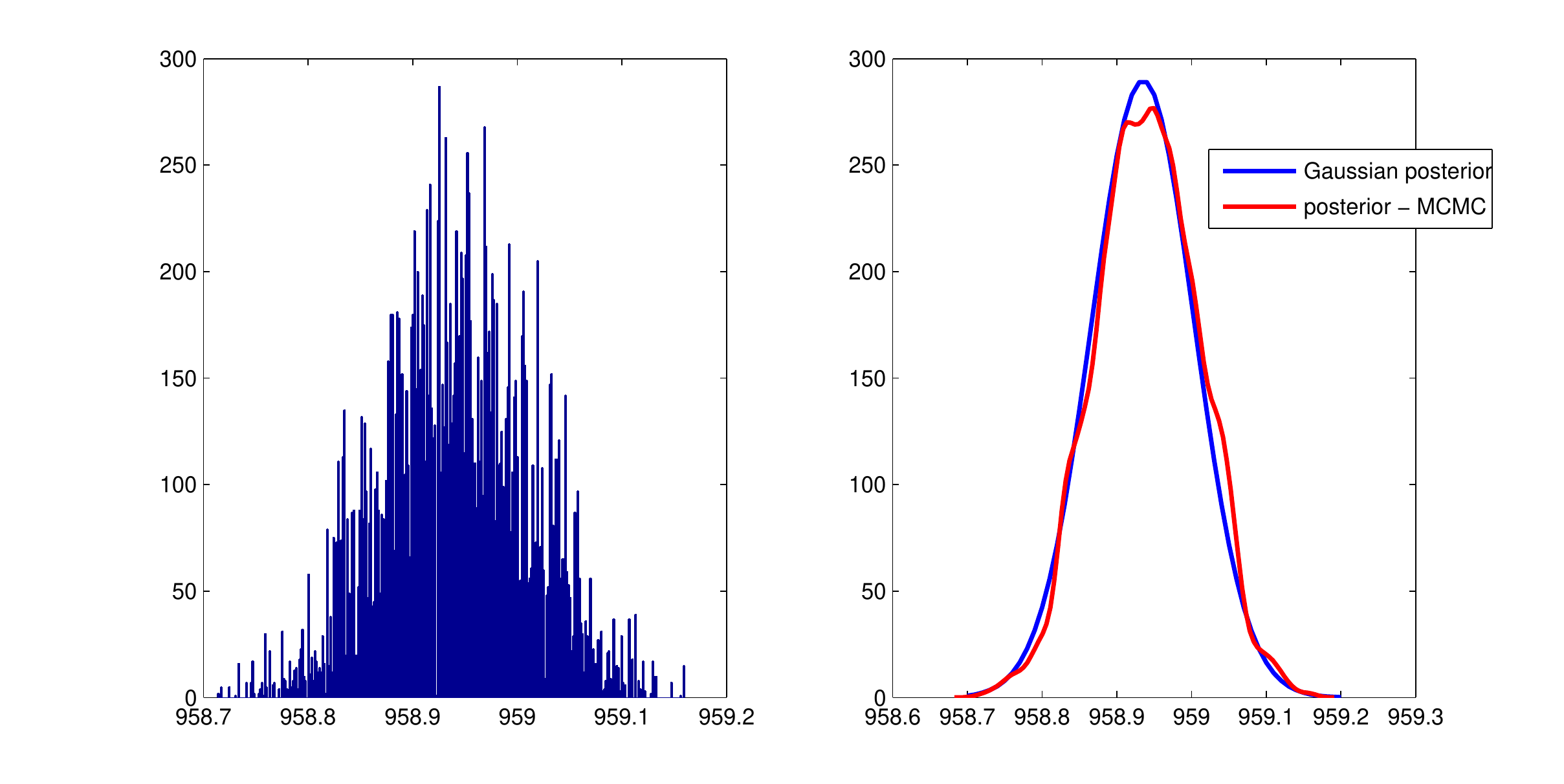}
\caption{Posterior for $a_{11}$}
\label{fig:posterior}
\end{figure}

\begin{figure}[tp]
\centering
\includegraphics[keepaspectratio=true, width = 5.5in]{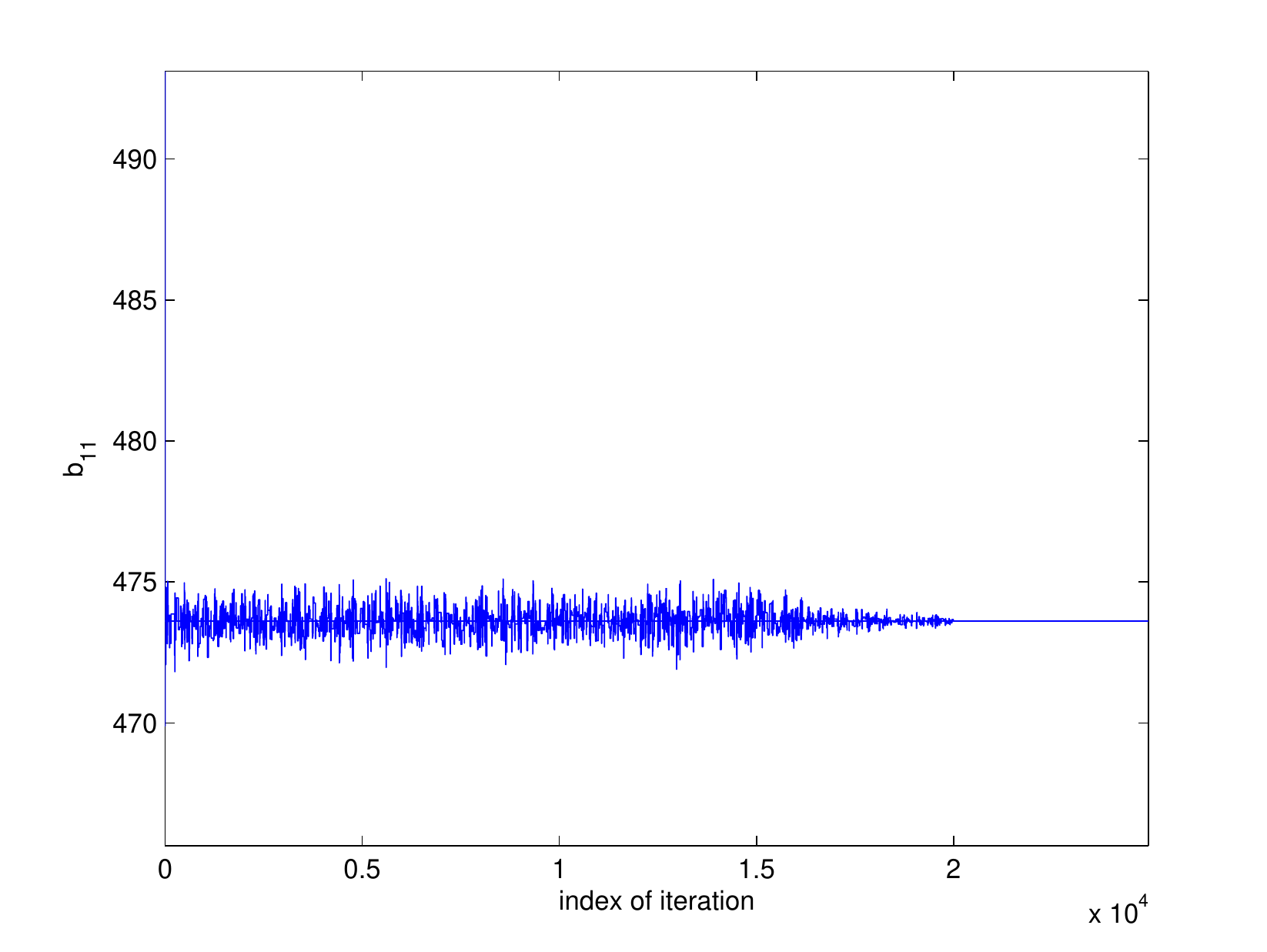}
\caption{Plot of the chain searching for $b_{11}$ in $presence$ of noise.}
\label{fig:chain_b11}
\end{figure}

This was implemented [Fig. \ref{fig:chain_a11} \& \ref{fig:chain_b11}], and in presence of noise too, the chains succeeded in converging at the true value. Even with inclusion of noise, the chains were observed to still converge extremely rapidly. Notable is, that, the starting value was always chosen randomly (using a random number generator to choose from a range of values, of course).

Also, the posterior was plotted for these two PN coefficients [Fig. \ref{fig:posterior}].
\pagebreak

Given, the success of Markov Chains in searching for $a_{11}$ and $b_{11}$, with all other coefficients taken as known; the next idea was to see how the chains fare in searching for $a_{10}$ and $b_{10}$ (individually), with no knowledge of the coefficients of 5.5PN$^{th}$ term. In other words, searching for $a_{10}$ \& $b_{10}$, with $a_{11} = 0.0$ and $b_{11} = 0.0$. In this, the attempt essentially was to take a waveform till lower PN approximation, and try to find the next PN coefficient, matching against a given higher Post-Newtonian approximation waveform.

\begin{figure}[tp]
\centering
\includegraphics[keepaspectratio=true, width = 5.5in]{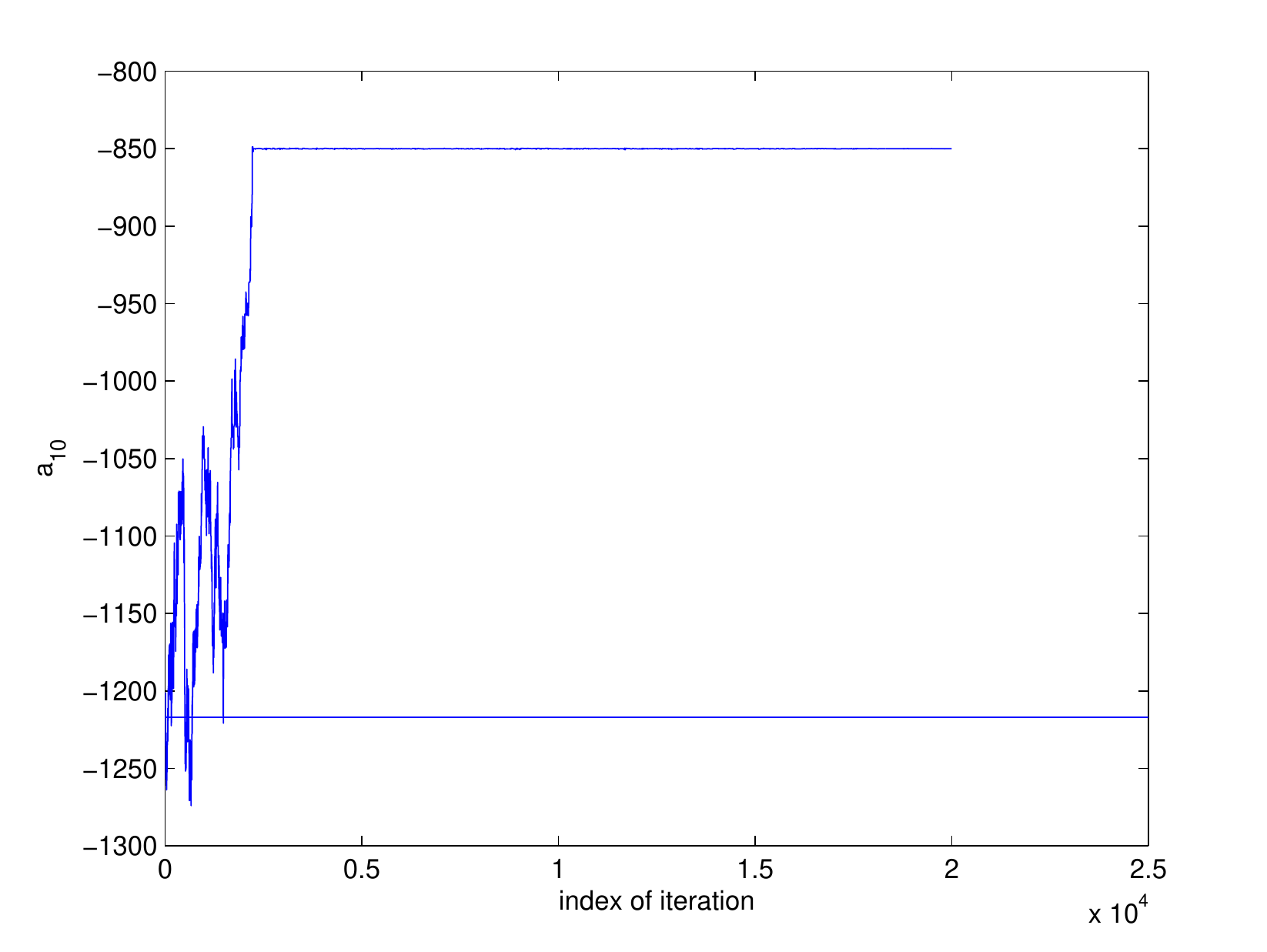}
\caption{Plot of the chain searching for $a_{10}$ in $presence$ of noise, and with $a_{11}=b_{11} =0$.}
\label{fig:chain_a10_20}

\centering
\includegraphics[keepaspectratio=true, width = 5.5in]{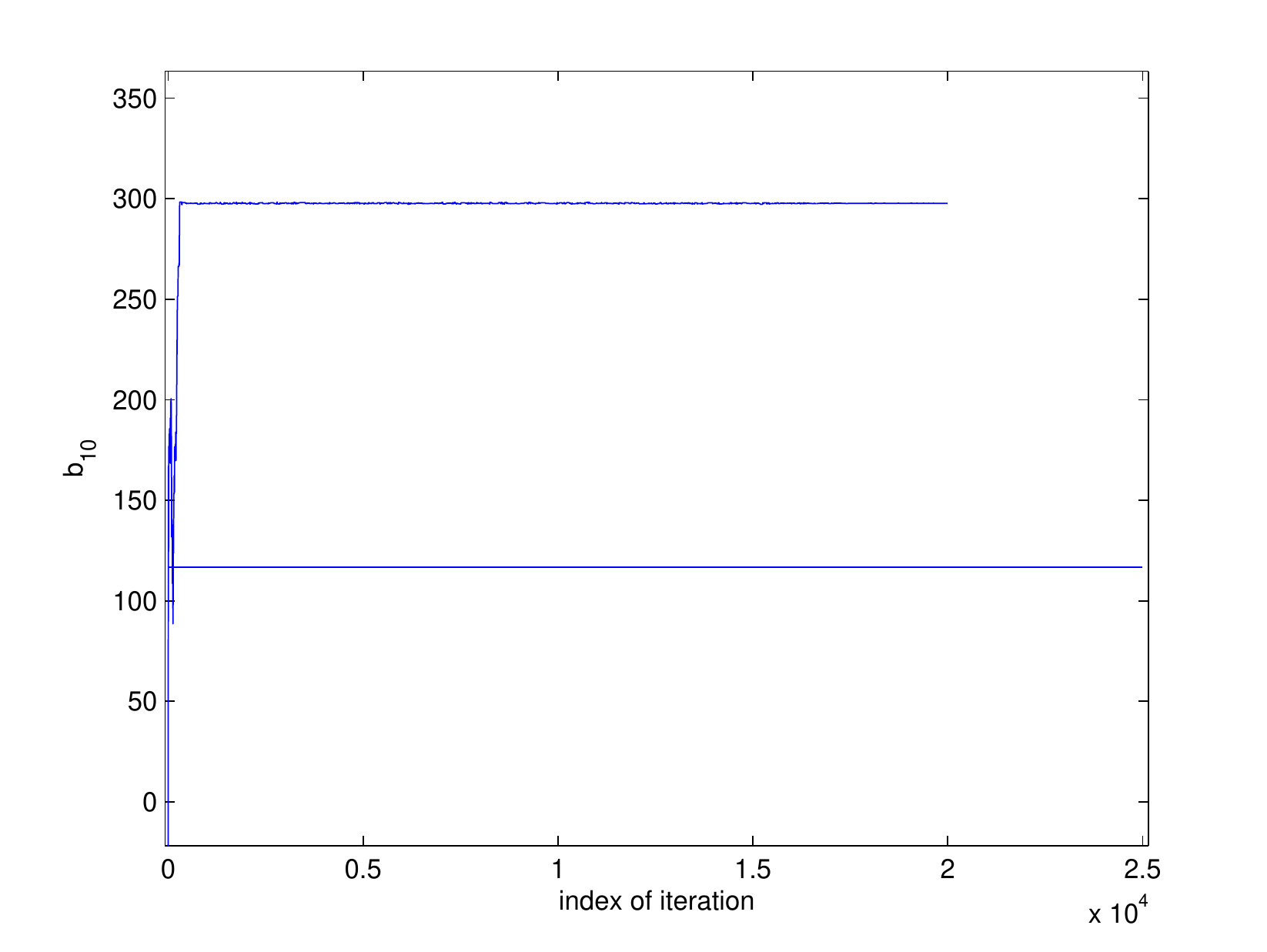}
\caption{Plot of the chain searching for $b_{10}$ in $presence$ of noise, and with $a_{11}=b_{11} =0$.}
\label{fig:chain_b10_20}
\end{figure}

However, this search did \textit{not} turn out positive results and several initial heats were tried out. The closest the chains gets, in the multitude of runs performed, is shown in Fig. \ref{fig:chain_a10_20} \& \ref{fig:chain_b10_20}. These searches were carried out with noisy original signal.

\begin{figure}[h]
\centering
\includegraphics[keepaspectratio=true,width=5.5in]{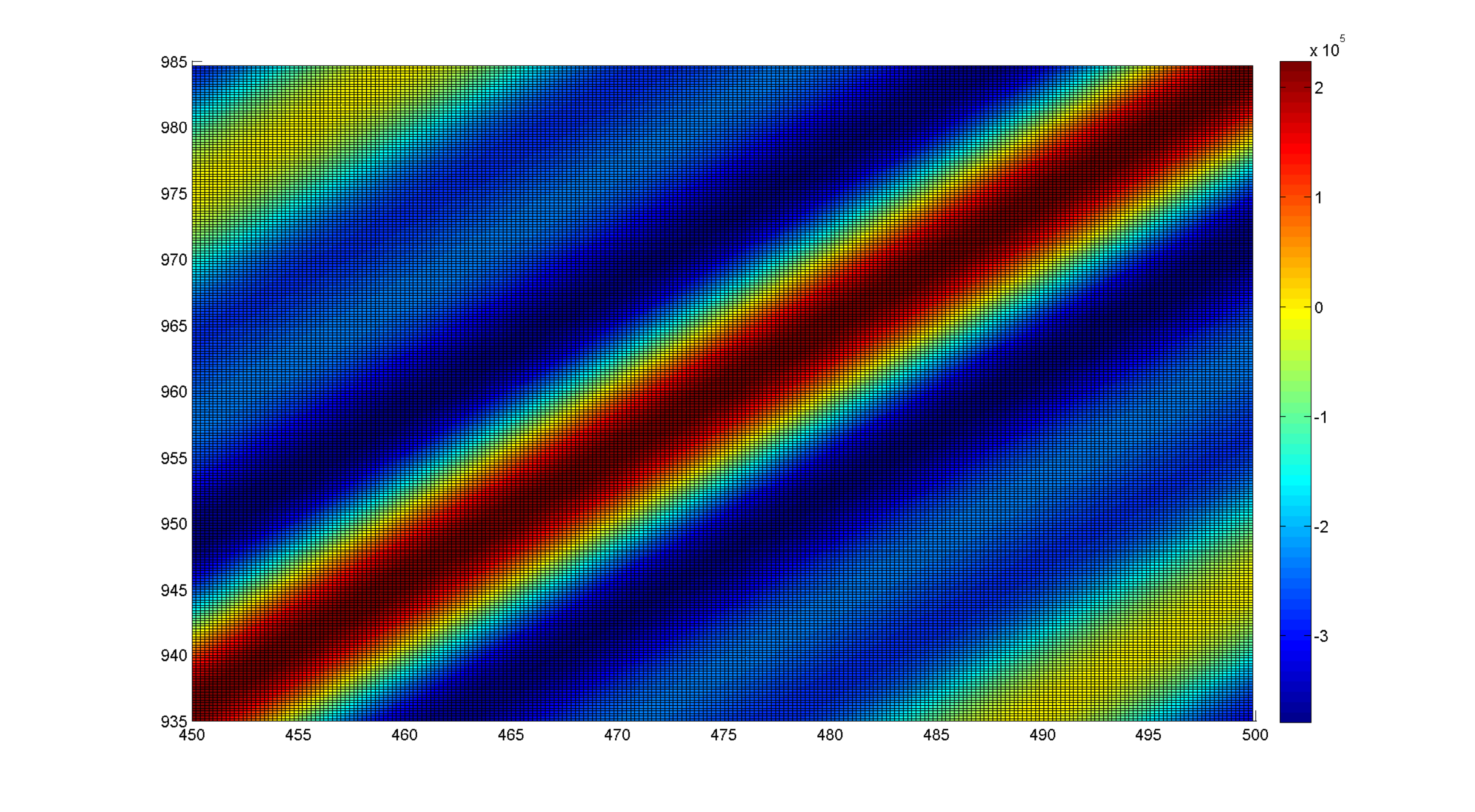}
\caption{LogLikelihood surface for the \{$a_{11},b_{11}$\} space, plotted as $a_{11} v/s b_{11}$.}
\label{fig:a11_b11}
\end{figure}

\pagebreak
The next step was to increase the dimensionality of the search space to 2, and search for $a_{11}$ and $b_{11}$ simultaneously (again keeping all the other coefficients of the 5.5PN expansion at their true values). Before that, the logLikelihood surface was plotted as a heat map for $a_{11}$ against $b_{11}$ [Fig. \ref{fig:a11_b11}].

With this, the Chains were set running to explore the 2-dimensional surface and search for the true values of $a_{11}$ \& $b_{11}$. The signal was initially taken with noise, and subsequently without it. The heat map [Fig. \ref{fig:a11_b11}] shows that there is a row of peaks running diagonal across the map, and Fig. \ref{fig:a11} \& \ref{fig:b11} were cross sections of this surface. This row of peaks means that it was going for the Chains to explore the surface completely and find the true maxima. The purpose of this search was to see, that if we knew the waveform to 5PN approximation, and the true signal be approximated by the 5.5PN waveform, then whether Markov Chains find the 5.5PN$^{th}$ coefficients.

\begin{figure}[tp]
\centering
\includegraphics[keepaspectratio=true,width=5.5in]{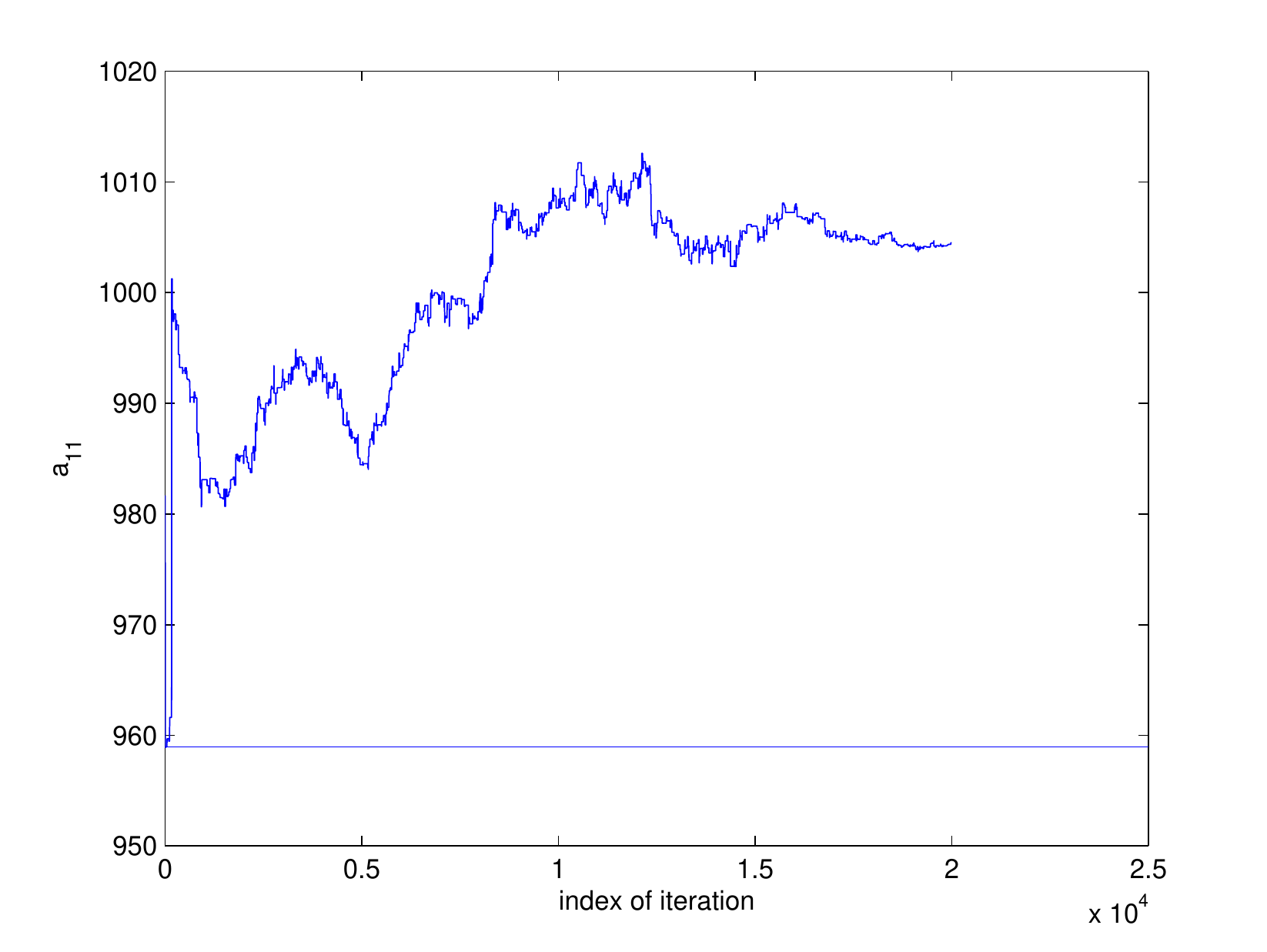}
\caption{Chain searching for $a_{11}$ (and $b_{11}$), eventually random-walking.}
\label{fig:chain_a11_2}

\centering
\includegraphics[keepaspectratio=true,width=5.5in]{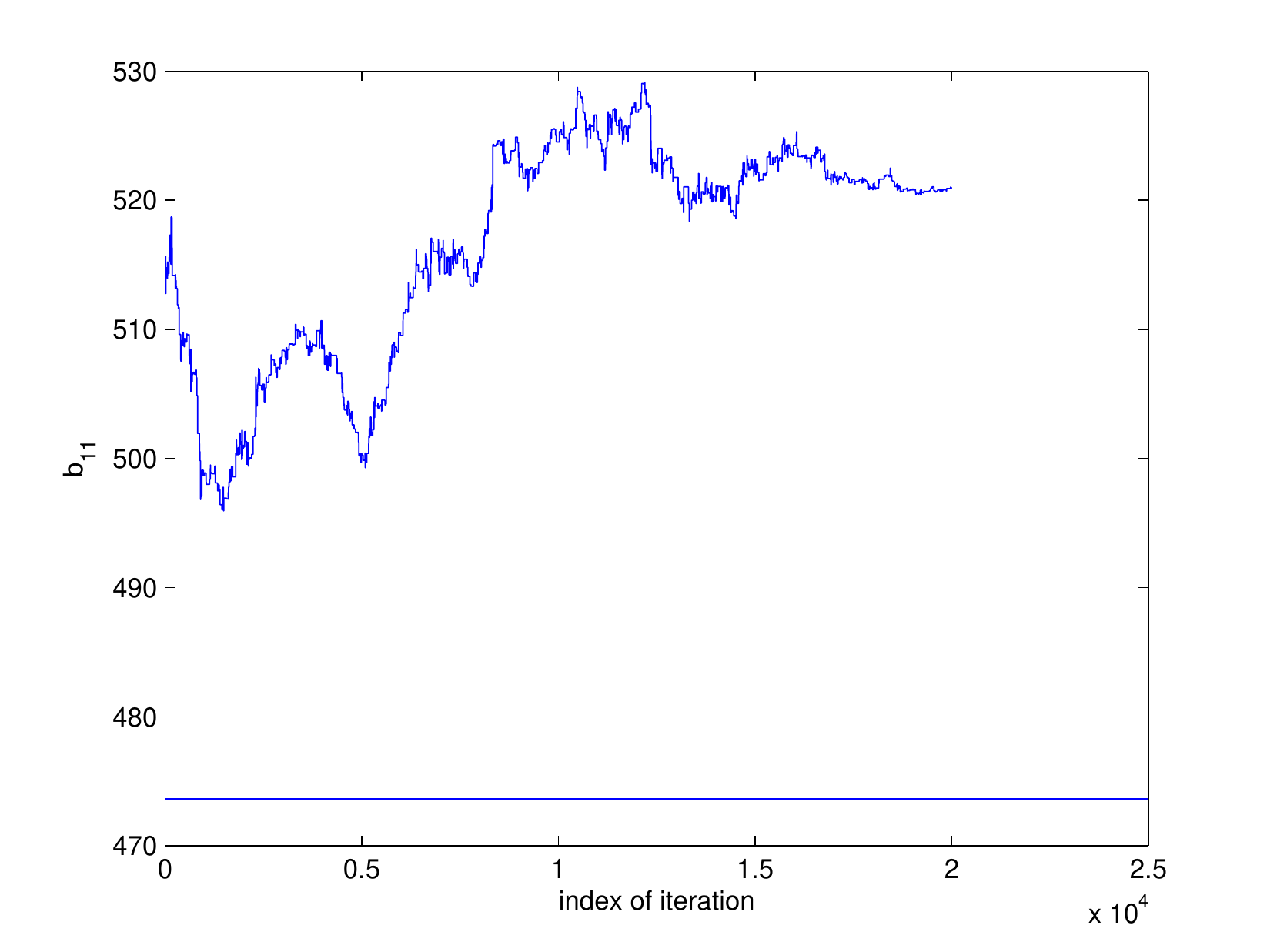}
\caption{Chain searching for $b_{11}$ (and $a_{11}$), eventually random-walking.}
\label{fig:chain_b11_2}
\end{figure}

Fig. \ref{fig:chain_a11_2} and \ref{fig:chain_b11_2}, show a very typical result for this case. All the chains set out, with different initial heats, and different initial values, ended up in random walk like this.\pagebreak

\begin{figure}[h]
\centering
\includegraphics[keepaspectratio=true,width=5.5in]{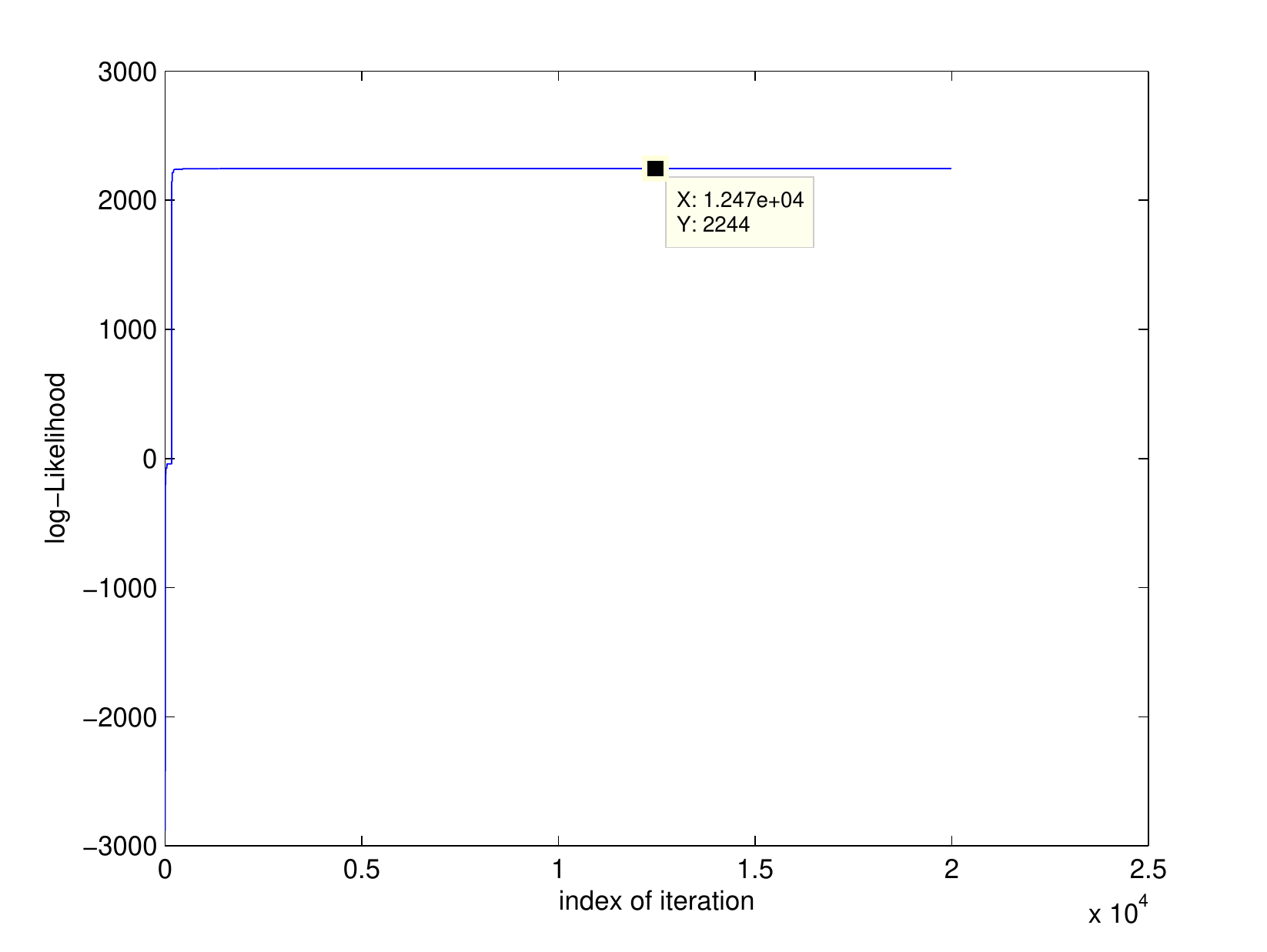}
\caption{LogLikelihood values for the Chain searching for $a_{11}$ \& $b_{11}$, using the true signal with noise.}
\label{fig:chain_logL_2}
\end{figure}

Important to note here [Fig. \ref{fig:chain_logL_2}], is the value of the local maxima it got stuck at. The value is $logL = 2.244 \times 10^3$, which differs only by $0.31\% $ from the logL value for a perfect match ($= 2.237016 \times 10^3$).

\begin{figure}[tp]
\centering
\includegraphics[keepaspectratio=true,width=5.5in]{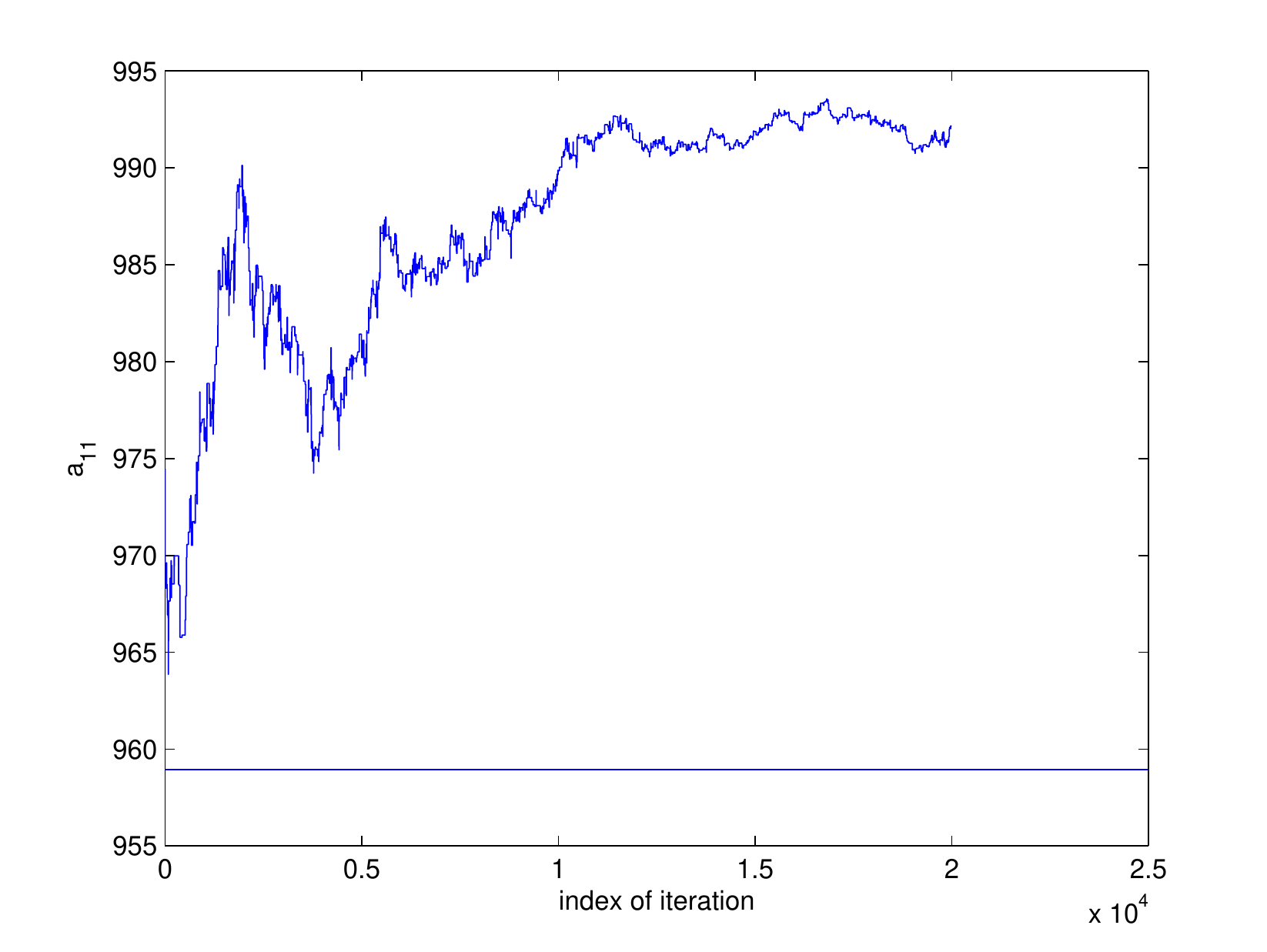}
\caption{Chain searching for $a_{11}$ (and $b_{11}$), eventually random-walking.The original signal used here is without noise.}
\label{fig:chain_a11_2WN}

\centering
\includegraphics[keepaspectratio=true,width=5.5in]{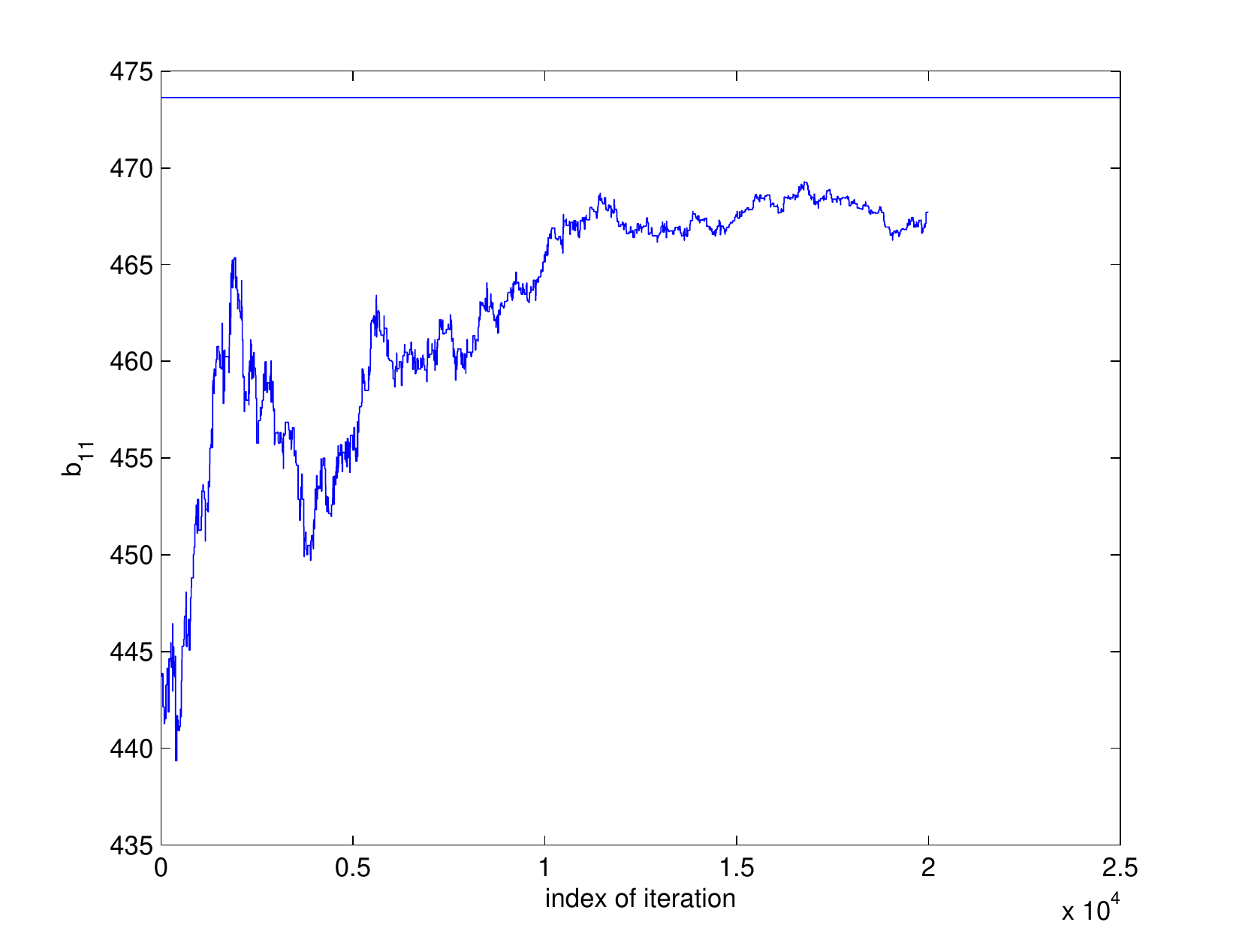}
\caption{Chain searching for $b_{11}$ (and $a_{11}$), eventually random-walking. The original signal used here is without noise.}
\label{fig:chain_b11_2WN}
\end{figure}

Another attempt at the same was made, with taking the signal without noise. The results are given in the figures \ref{fig:chain_a11_2WN} and \ref{fig:chain_b11_2WN}. Not surprisingly, the results did not improve, and the chain repeated its behaviour by wandering away from the true value.

\pagebreak

Further, search chains were also setup for searching for coefficients in 3 dimensions. Which means, $m = 2 \& n = 1$ [\ref{eq:defmn}]. The true signal was taken as the 5.5Post Newtonian approximation, without noise. These chains showed the following result:

\begin{figure}[h]
\centering
\includegraphics[keepaspectratio=true,width=5.5in]{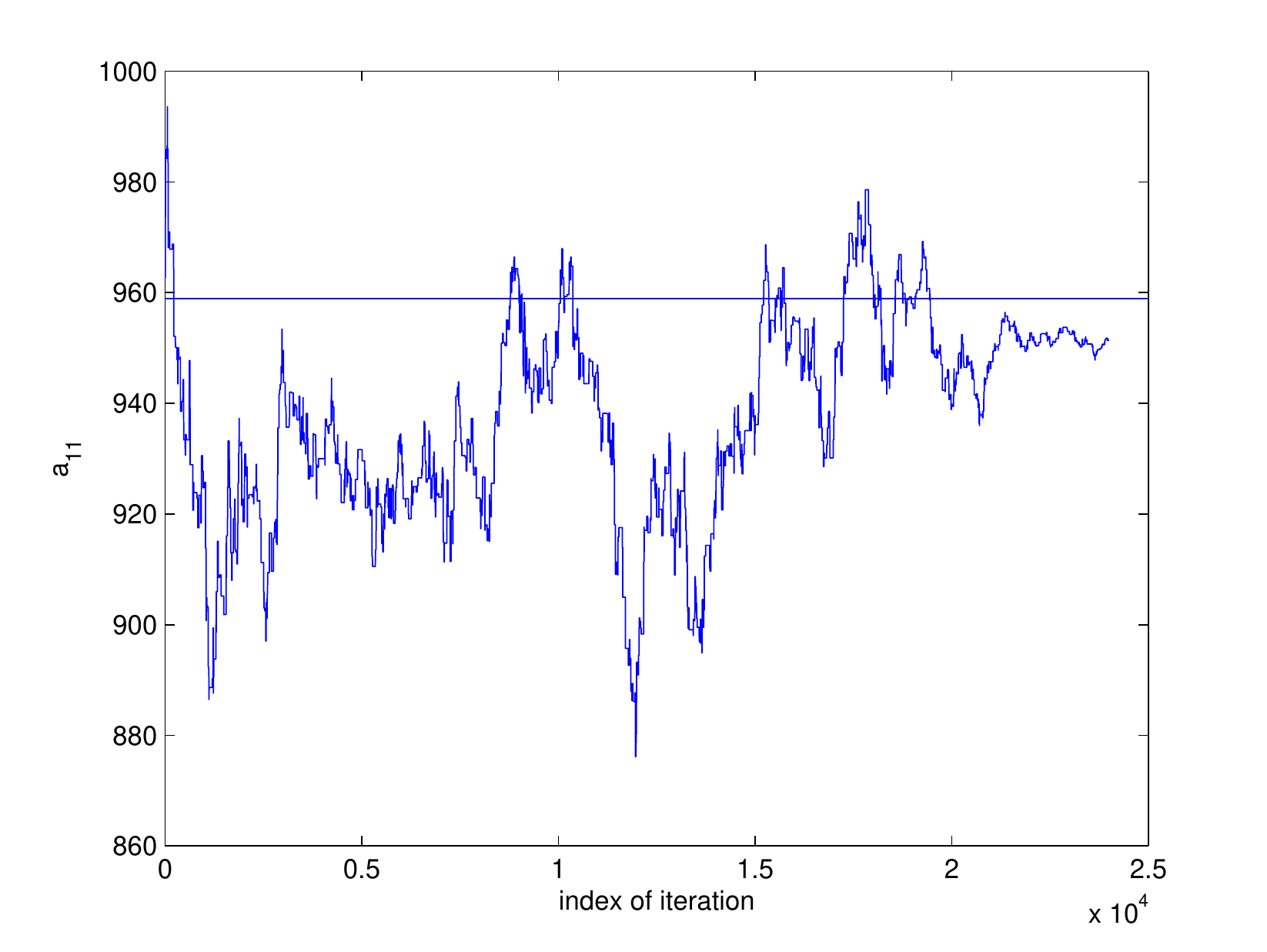}
\caption{Chain searching for $a_{11}$ (\& $b_{11}, a_{10}$), using the true signal without noise.}
\label{fig:chain_a11_3}
\end{figure}

\begin{figure}[tp]
\centering
\includegraphics[keepaspectratio=true,width=5.5in]{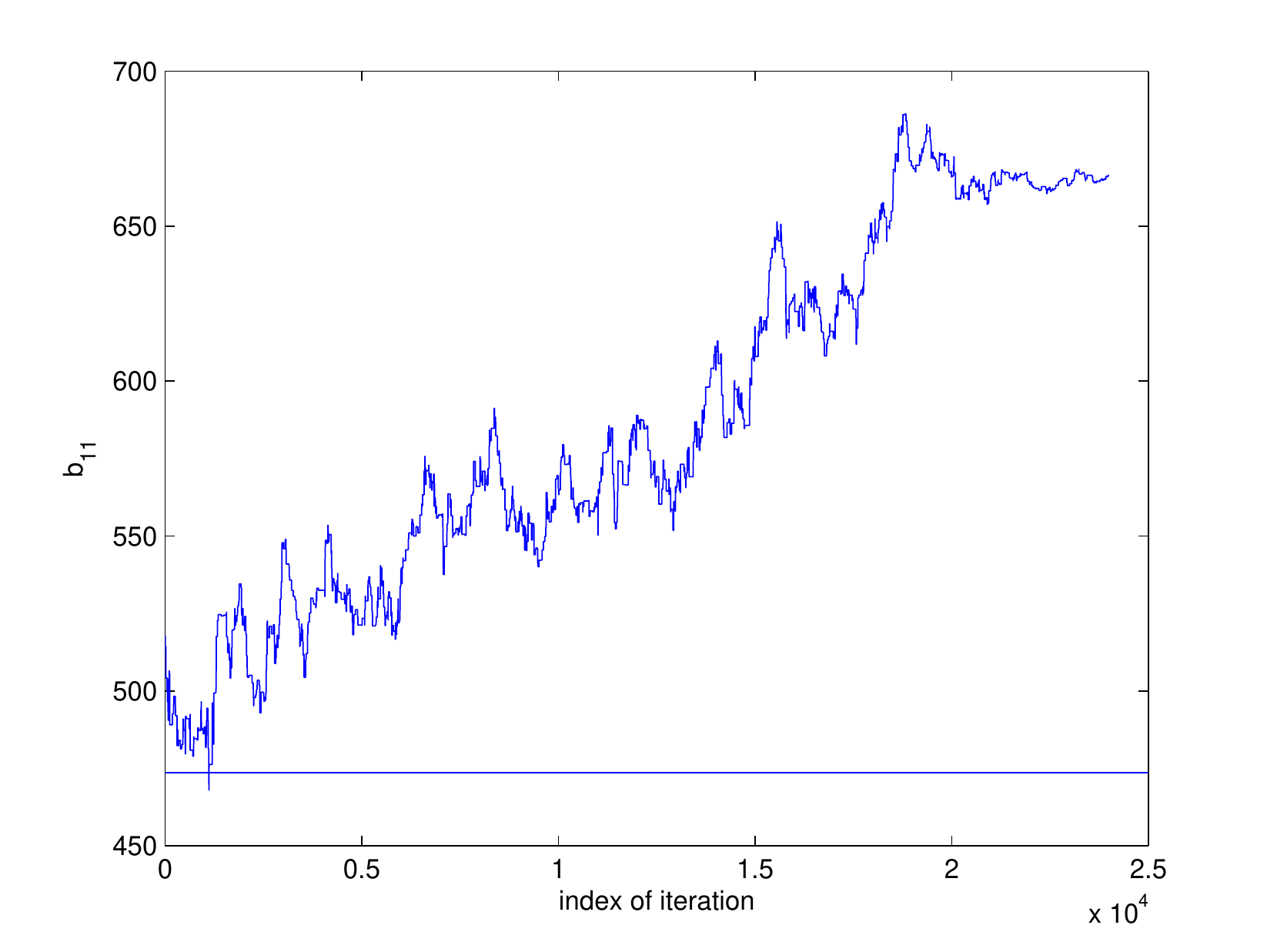}
\caption{Chain searching for $b_{11}$ (\& $a_{11}, a_{10}$), using the true signal without noise.}
\label{fig:chain_b11_3}

\centering
\includegraphics[keepaspectratio=true,width=5.5in]{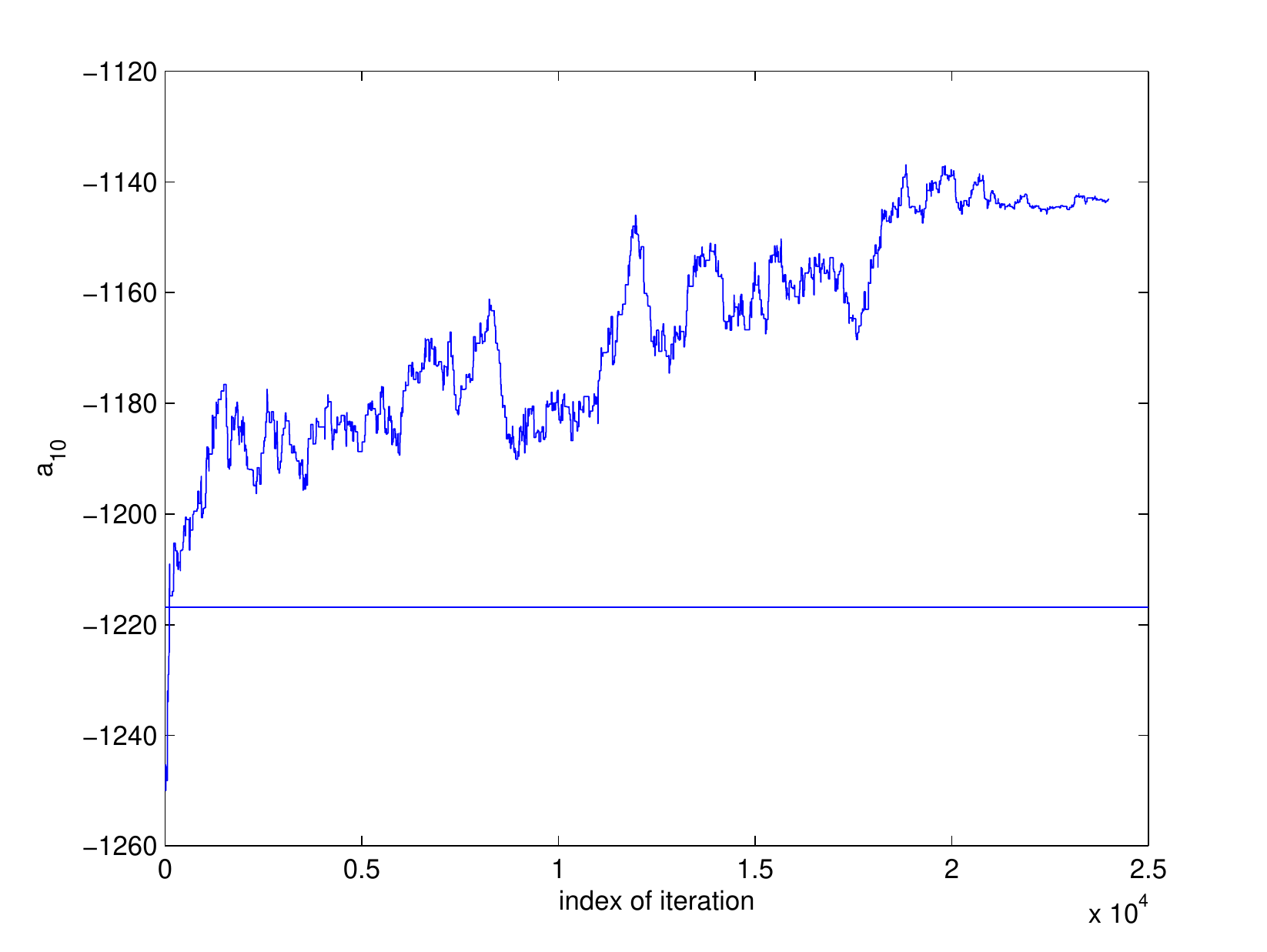}
\caption{Chain searching for $a_{10}$ (\& $a_{11}, b_{11}$), using the true signal without noise.}
\label{fig:chain_a10_3}
\end{figure}

\chapter{Conclusion}
In this thesis, the main idea dealt with is to search for post-Newtonian coefficients, for the gravitational flux radiated from Massive Black Hole Binary Inspirals - in the test mass limit. The search was done using stochastic search, employing the Markov Chain Monte Carlo method\cite{Metro_Hastings} \cite{Simulated_Annealing1}. The idea germinated from the fact that to come up with more accurate expressions for the gravitational flux, higher order post - Newtonian approximation would be required. And MCMC is a promising way to go about it.

Also, when worked out against actual data from LISA, the coefficients that the Markov Chains come up with can be compared with those predicted by the post - Newtonian theory. This would actually work out as a test for General Relativity, and a verification of its predictions. The approach was tried against existing 5.5 order post - Newtonian expression with the mathematical model for the LISA noise\cite{instr_noise_LISA} \cite{galactic_noise1} \cite{galactic_noise2}, and the chains managed to find the last coefficients extremely rapidly. Also mentionable is that the values they settled at, were within $1/500^{th}$ of a percent of the true values. This is illustrated in the folowing histogram plots, for $a_{11}$ and $b_{11}$ [Eq. \ref{eq:g_flux}].
\begin{figure}[tp]
\centering
\includegraphics[keepaspectratio=true, width = 6.0in]{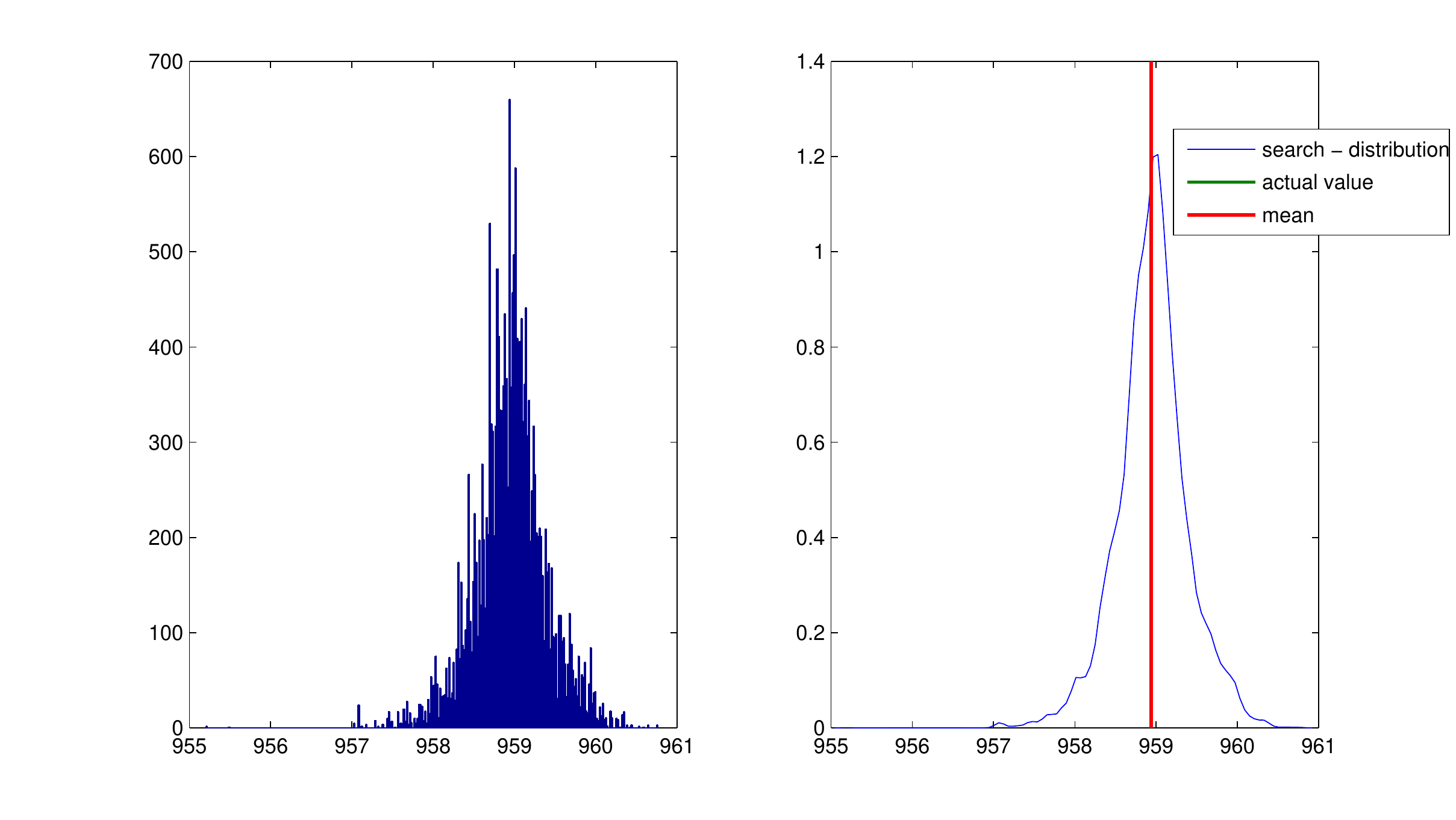}
\caption{Distribution of the values taken by the chain for $a_{11}$ in $presence$ of noise.}
\label{fig:distribution_a11}

\centering
\includegraphics[keepaspectratio=true, width = 5.5in]{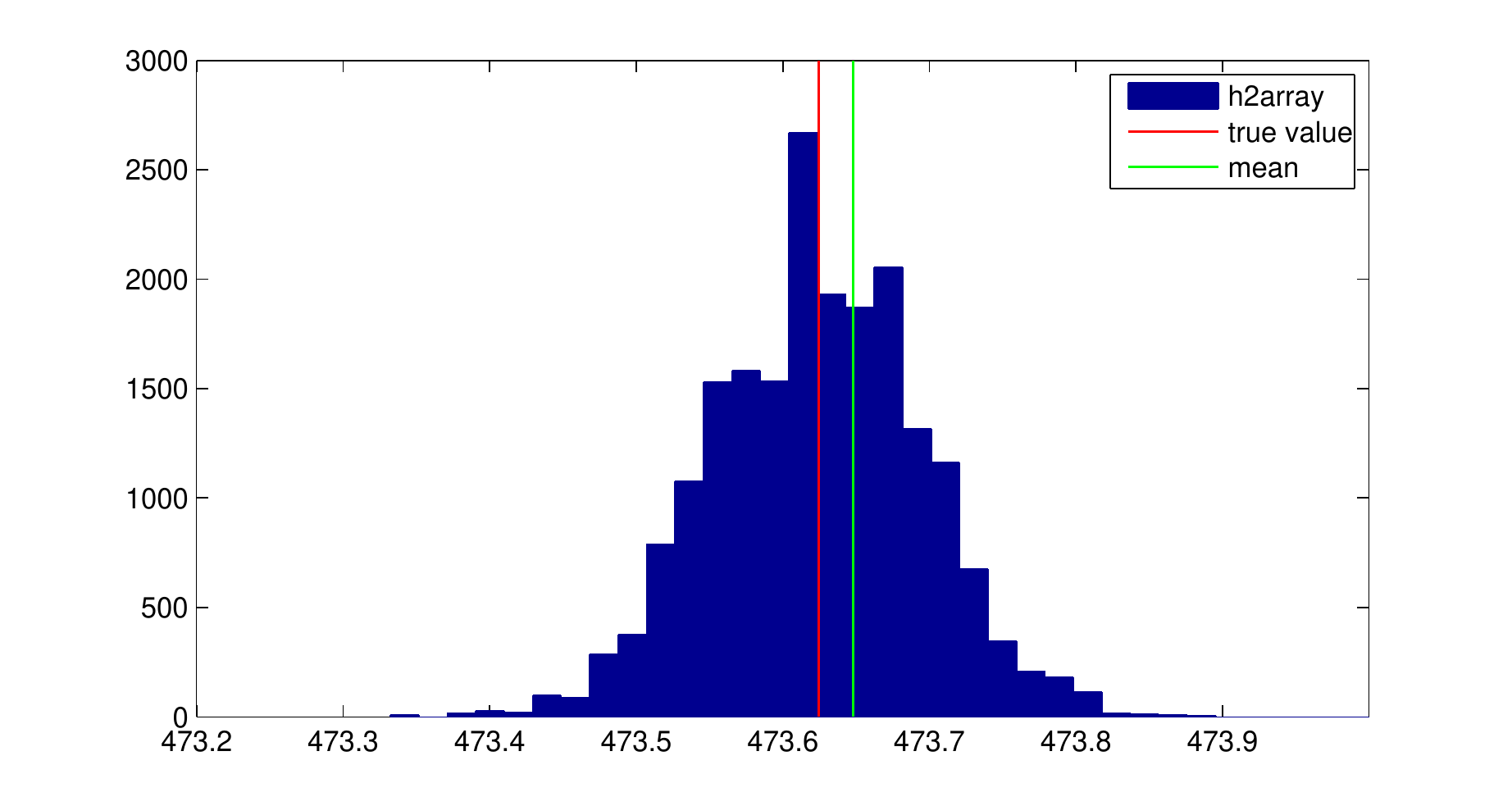}
\caption{Distribution of the values taken by the chain for $b_{11}$ in $presence$ of noise.}
\label{fig:distribution_b11}
\end{figure}

\pagebreak However, the searches attempted beyond trying to search for the last coefficient, did not yield positive results. Several cooling schedules [Sec.\ref{subsec:approach}] and starting temperatures were experimented with. However, the chains failed to yield persistent results. Theoretically pondering, this should have worked, but it did not. On investigating for the reason, it was found in the plenitude of local maxima found in the neighborhood of the global maxima. This is clearly illustrated in a surface plot of the log-Likelihood [Fig.\ref{fig:3d_a11_b11}] (which was used as the indicator to the match of the signal evolved using the predicted coefficients and the ideal signal) against the last two coefficient - dimensions:
\begin{figure}[h]
\centering
\includegraphics[keepaspectratio=true, width = 6.0in]{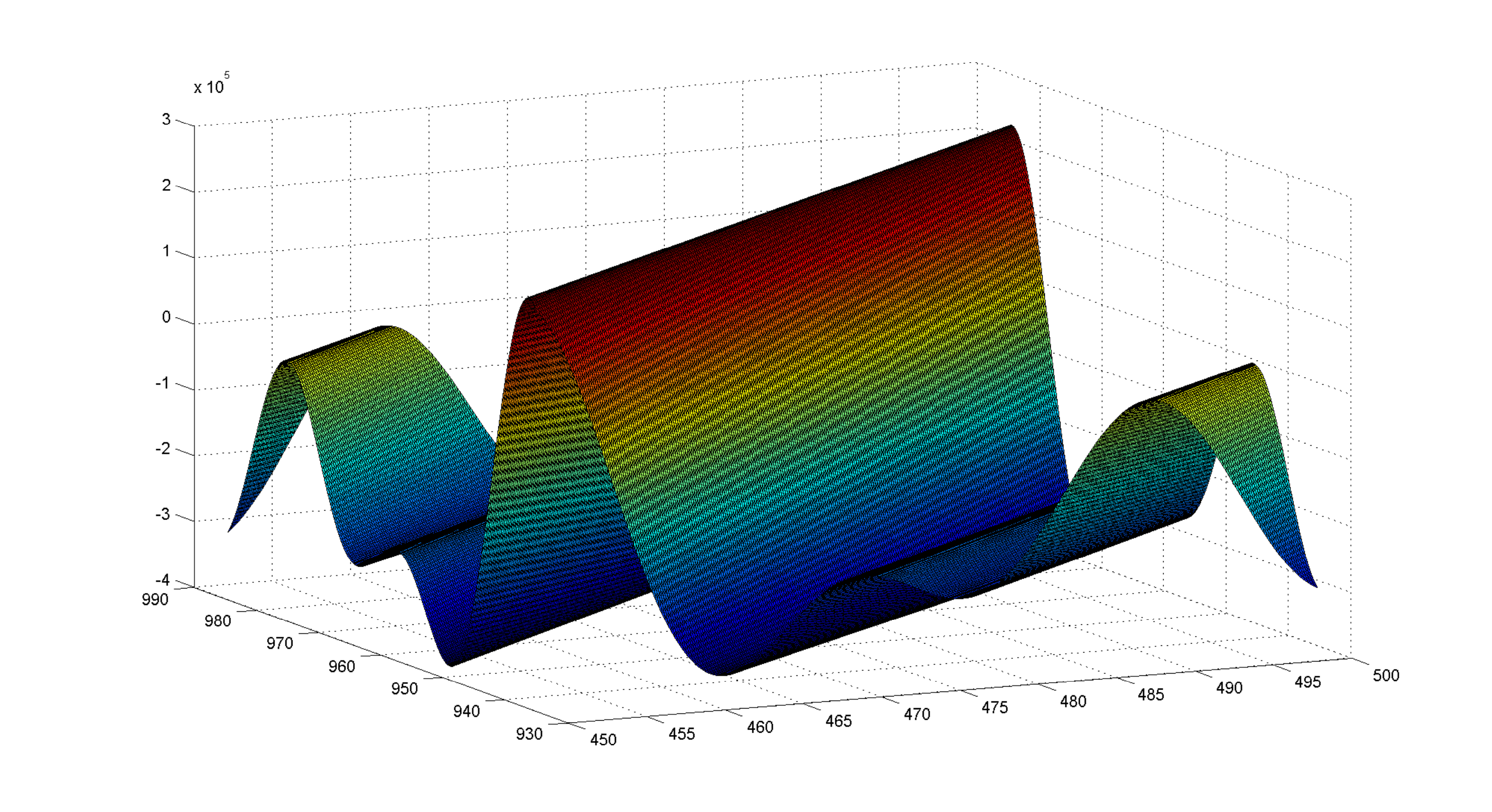}
\caption{Log-Likelihood surface, in the space of the last two coefficients ($a_{11}$ and $b_{11}$).}
\label{fig:3d_a11_b11}
\end{figure}

It is sufficiently clear that there is a row of maxima running across the plane of the two dimensions. The chain kept getting stuck on these maxima, which were found to be as close as within 0.25 \% of the true perfect-match log-Likelihood. However this is to be treated as a preliminary study of this approach and in future, the work is hoped to be extended to searching for physical parameters along-with the post-Newtonian coefficients. It would be even more interesting to see if this approach works when the physical parameters of the system are not fixed themselves (as they were, throughout here). The idea is, to have MCMC to search for the parameters like the sky position or the masses, after an initial preliminary search and narrowing down of the candidates has been accomplished, along-with searching for the Post-Newtonian coefficients. The coefficients found can be matched against the theoretically predicted values, and the comparison can be well treated as a test for the PN theory, and that of GR itself.

\newpage

\bibliographystyle{plain}
\bibliography{thesis}

\newpage
\end{document}